\documentclass[twocolumn]{aastex631}
\usepackage{float}

\def\arcsec{\hbox{\ensuremath{^{\prime\prime}}}}
\def\farcs{\hbox{\ensuremath{.\!\!^{\prime\prime}}}}
\newcommand{\kms}{km\,s\ensuremath{^{-1}}}
\newcommand{\mbh}{\ensuremath{M_\mathrm{BH}}}

\def\farcs{\hbox{\ensuremath{.\!\!^{\prime\prime}}}}

\newcommand{\sigmastar}{\ensuremath{\sigma_\star}}
\newcommand{\chinu}{\ensuremath{\chi^2_{\mathrm{\nu}}}}

\graphicspath{{./}{figures/}}

\begin{document}

\title{Gas-dynamical Mass Measurements of the Supermassive Black Holes in the Early-Type Galaxies NGC 4786 and NGC 5193 from ALMA and HST Observations\footnote{Based on observations made with the NASA/ESA Hubble Space Telescope, obtained at the Space Telescope Science Institute, which is operated by the Association of Universities for Research in Astronomy, Inc., under NASA contract NAS5-26555. These observations are associated with programs 15226 and 15909.}}

\author[0000-0003-2632-8875]{Kyle M. Kabasares}
\affiliation{Ames Research Center, National Aeronautics and Space Administration, Moffett Field, CA 94035, USA}
\affiliation{Bay Area Environmental Research Institute, Ames Research Center, Moffett Field, CA 94035, USA}
\affiliation{Department of Physics and Astronomy, 4129 Frederick Reines Hall, University of California, Irvine, CA, 92697-4575, USA}

\author[0000-0003-1420-6037]{Jonathan H. Cohn}
\affiliation{Department of Physics and Astronomy, Dartmouth College, 6127 Wilder Laboratory, Hanover, NH 03755, USA}
\affiliation{George P. and Cynthia Woods Mitchell Institute for Fundamental Physics and Astronomy, 4242 TAMU, Texas A\&M University, College Station, TX, 77843-4242,}

\author[0000-0002-3026-0562]{Aaron J. Barth}
\affiliation{Department of Physics and Astronomy, 4129 Frederick Reines Hall, University of California, Irvine, CA, 92697-4575, USA}

\author[0000-0001-6301-570X]{Benjamin D. Boizelle}
\affil{Department of Physics and Astronomy, 284 ESC, Brigham Young University, Provo, UT, 84602, USA}

\author[0000-0003-3900-6189]{Jared Davidson}
\affil{Department of Physics and Astronomy, 284 ESC, Brigham Young University, Provo, UT, 84602, USA}

\author{Janelle M. Sy}
\affiliation{Department of Physics, New York University, 726 Broadway, New York, NY, 10003, USA}
\affiliation{Department of Physics and Astronomy, 4129 Frederick Reines Hall, University of California, Irvine, CA, 92697-4575, USA}

\author[0000-0002-7346-9868]{Jeysen Flores-Vel\'azquez}
\affiliation{Institute for Gravitation and the Cosmos, The Pennsylvania State University, University Park, PA 16802, USA}
\affiliation{Department of Physics and Astronomy, 4129 Frederick Reines Hall, University of California, Irvine, CA, 92697-4575, USA}

\author[0000-0003-3940-0899]{Silvana C. Delgado Andrade}
\affiliation{George P. and Cynthia Woods Mitchell Institute for Fundamental Physics and Astronomy, 4242 TAMU, Texas A\&M University, College Station, TX, 77843-4242,}

\author[0000-0002-3202-9487]{David A. Buote}
\affiliation{Department of Physics and Astronomy, 4129 Frederick Reines Hall, University of California, Irvine, CA, 92697-4575, USA}

\author[0000-0002-1881-5908]{Jonelle L. Walsh}
\affiliation{George P. and Cynthia Woods Mitchell Institute for Fundamental Physics and Astronomy, 4242 TAMU, Texas A\&M University, College Station, TX, 77843-4242,}

\author[0000-0002-7892-396X]{Andrew J. Baker}
\affiliation{Department of Physics and Astronomy, Rutgers, the State University of New Jersey, 136 Frelinghuysen Road, Piscataway, NJ 08854-8019, USA}
\affiliation{Department of Physics and Astronomy, University of the Western Cape, Robert Sobukwe Road, Bellville 7535, South Africa}

\author[0000-0003-2511-2060]{Jeremy Darling}
\affiliation{Center for Astrophysics and Space Astronomy, Department of Astrophysical and Planetary Sciences, University of Colorado, 389 UCB, Boulder, CO 80309-
0389, USA}

\author[0000-0001-6947-5846]{Luis C. Ho}
\affiliation{Kavli Institute for Astronomy and Astrophysics, Peking University, Beijing 100871, China; Department of Astronomy, School of Physics, Peking University,
Beijing 100871, People's Republic of China}

\correspondingauthor{Kyle M. Kabasares}
\email{kabasar@baeri.org}
\footnote{This study was jointly led by Dr. Kyle M. Kabasares and Dr. Jonathan H. Cohn.}



\begin{abstract}
We present molecular gas-dynamical mass measurements of the central black holes in the giant elliptical galaxies NGC 4786 and NGC 5193, based on CO(2$-$1) observations from the Atacama Large Millimeter/submillimeter Array (ALMA) and Hubble Space Telescope near-infrared imaging. The central region in each galaxy contains a circumnuclear disk that exhibits orderly rotation with projected line-of-sight velocities of ${\sim} 270\, \mathrm{km}\,\mathrm{s^{-1}}$. We build gas-dynamical models for the rotating disk in each galaxy and fit them directly to the ALMA data cubes. At $0\farcs{31}$ resolution, the ALMA observations do not fully resolve the black hole sphere of influence (SOI), and neither galaxy exhibits a central rise in rotation speed, indicating that emission from deep within the SOI is not detected. As a result, our models do not tightly constrain the central black hole mass in either galaxy, but they prefer the presence of a central massive object in both galaxies. We measure the black hole mass to be $(\mbh/10^8\, M_{\odot}) = 5.0 \pm 0.2 \,[\mathrm{1\sigma \,statistical}] \,^{+1.4}_{-1.3} \,[\mathrm{systematic}]$ in NGC 4786 and $(\mbh/10^8\, M_{\odot}) = 1.4 \pm 0.03 \, [\mathrm{1\sigma\,statistical}] ^{+1.5}_{-0.1} \,[\mathrm{systematic}]$ in NGC 5193. The largest component of each measurement's error budget is from the systematic uncertainty associated with the extinction correction in the host galaxy models. This underscores the importance of assessing the impact of dust attenuation on the inferred $\mbh$.

\end{abstract}



\section{Introduction}
Supermassive black holes (BHs) are thought to reside at the centers of most, if not all, massive galaxies. With masses between a million to over a billion times that of the Sun, supermassive BHs gravitationally dominate the orbits of objects within their sphere of influence (SOI). The radius of the SOI is often defined as either $r_{\mathrm{SOI}} \approx G\mbh/\sigmastar^2$, where $\sigmastar$ represents the stellar velocity dispersion of the spheroidal component of the galaxy, or the radius where the enclosed galaxy and BH mass are equal $M_{\star}(r_{\mathrm{SOI}}) = \mbh $. While the SOI of a supermassive BH is limited to the innermost part of a galaxy, research has identified scaling relations between the mass of the central supermassive BH and large-scale properties of the host galaxy such as the spheroidal component's stellar velocity dispersion, luminosity, and mass. These relations suggest that galaxies and their supermassive BHs coevolve through cosmic time and regulate each other's growth \citep{2009ApJ...698..198G,2013ARAA..51..511K,2013ApJ...764..184M,2016ApJ...818...47S}. 

Increasing the sample of reliably measured BH masses is vital to understanding the scaling relations and BH-host galaxy coevolution as a whole. Over the last three decades, there have been approximately 100 supermassive BH mass measurements obtained primarily through either stellar or ionized gas-dynamical modeling \citep{2013ARAA..51..511K}. A key factor in securing a robust dynamical BH mass measurement is using observations that can probe scales near or within the projected radius of the BH's SOI. For extragalactic sources, $\mathrm{H}_2\mathrm{O}$ megamaser emission can be observed at very high angular resolution through Very Long Baseline Interferometry and have been used to secure some of the most precise extragalactic BH masses to date \citep{1995Natur.373..127M,2011ApJ...727...20K,2018ApJ...859..172K}, but given that they are very rare \citep{2016ApJ...819...11V}, other tracers must be used to understand BH demographics.

Since it became operational within the past decade, the Atacama Large Millimeter/submillimeter Array (ALMA) has been used to observe the rotation of molecular gas disks to constrain BH masses in nearby galaxies. In the best cases, ALMA has provided BH mass measurements with uncertainties of ${\leq} 10\%$ \citep{2016ApJ...822L..28B,2019ApJ...881...10B,2019MNRAS.490..319N}. In this paper, we add to the growing number of BH mass measurements in early-type galaxies (ETGs) with ALMA (\citealp{2016ApJ...822L..28B,2017MNRAS.468.4675D,2017MNRAS.468.4663O,2018MNRAS.473.3818D,2019MNRAS.485.4359S,2021MNRAS.503.5984S,2021ApJ...908...19B,2021ApJ...919...77C,2022ApJ...934..162K,2023arXiv230406117R}) by dynamically measuring the masses of the central BHs in the ETGs NGC 4786 and NGC 5193. 

These two galaxies were selected as part of an ALMA program designed to identify rapidly rotating molecular gas on scales comparable to $r_{\mathrm{SOI}}$. The selection was based on the identification of a smooth and morphologically round circumnuclear dust disk from optical Hubble Space Telescope (HST) images. ALMA CO(2$-$1) observations presented by \cite{PhD2018Boizelle} revealed that their molecular gas kinematics are dominated by disklike rotation, which we use to constrain the masses of their central BHs. The optical dust disks in both galaxies are relatively small. In angular size, the dust disk in NGC 4786 has a radius of about $0\farcs{6}$ or 181 pc and the disk in NGC 5193 has a radius of about $1\arcsec$ or 221 pc given our assumed angular diameter distances (discussed further in this section). The projected line-of-sight (LOS) velocities of the molecular gas in both disks exhibit similar features as they are in excess of ${\sim} 270\, \mathrm{km}\,\mathrm{s^{-1}}$ in the outer parts of the disk and remain somewhat flat at radii extending toward the disk center.

NGC 4786 is classified as a cD pec galaxy in the Third Reference Catalog of Bright Galaxies (RC3; \citealp{1991rc3..book.....D}). Redshift-independent distances for this galaxy in the NASA/IPAC Extragalactic Database \footnote{https://ned.ipac.caltech.edu}{(NED)} are between $65 - 75$ Mpc when using a $\Lambda$CDM cosmology with $H_0 = 67.8\,\mathrm{\kms} \,\mathrm{Mpc}^{-1}$. We adopt a Hubble-flow distance, using a value of $H_0 = 73 \,\mathrm{\kms}\, \mathrm{Mpc}^{-1}$ based on more recent estimates of $H_0$ from nearby ($<100$ Mpc) galaxies \citep{2021ApJ...911...65B,2022ApJ...934L...7R,2022ApJ...935...83K}, a recessional velocity of $cz = 4623 \, \mathrm{\kms} $ from preliminary gas-dynamical models, $\Omega_M = 0.31$, and $\Omega_{\mathrm{\Lambda}} = 0.69$. These assumptions set a luminosity distance of 64.1 Mpc, an angular diameter distance of 62.1 Mpc, and an angular scale where $1\arcsec{}$ corresponds to 301 pc. The measured $\mbh$ scales linearly with the assumed distance, so any differences in assumed distance to the galaxy will correspond to an equivalent rescaling of the measured $\mbh$. There are no previous studies that have measured the mass of the supermassive BH in this galaxy.

NGC 5193 is classified as an E pec galaxy by RC3. In past literature, this galaxy has been associated with and sometimes identified as the brightest cluster galaxy in the Abell 3560 \citep{1989ApJS...70....1A} cluster with a recessional velocity  of $cz \sim 3800 - 4000\,\mathrm{\kms} $ \citep{1998ApJ...499..577L,2002ApJ...567..294O}. However, \cite{1990AJ.....99.1709V} and \cite{1999AJ....118.1131W} instead associate NGC 5193 with the Abell 3565 cluster based on findings from \cite{1987RMxAA..14...72M} that indicate Abell 3560 has a recessional velocity of $cz \sim$ 14,850$ \,\mathrm{\kms}$. This implies that Abell 3560 is a background galaxy cluster and that NGC 5193 is not a member.
\cite{2001ApJ...546..681T} determined a distance modulus of $m - M 
= 32.66 \pm 0.29$ mag for NGC 5193 with ground-based $I$-band surface brightness fluctuation (SBF) data while distinguishing it as separate from Abell 3560. This distance modulus translates to a luminosity distance of $34.0 \pm 4.5 $ Mpc, but they list this measurement as uncertain. \cite{2003ApJ...583..712J} independently measured a distance modulus of $m - M = 33.35 \pm 0.15$ mag, using SBF measurements from HST NICMOS F160W data. The corresponding luminosity distance is $46.8 \pm 3.2$ Mpc, and we adopt this distance for our dynamical modeling purposes. Using a recessional velocity of $cz = 3705 \,\mathrm{\kms}$ from initial gas-dynamical models, this gives an angular diameter distance of 45.7 Mpc, where $1\arcsec{}$ corresponds to 221 pc. As in the case of NGC 4786, there are no previous works that have constrained the central BH mass in this galaxy. 

We organize our paper as follows. In Section \ref{sec:Observations}, we describe our ALMA and HST observations as well as the data calibration and reduction process. In Section \ref{sec:HostGalaxyModeling}, we model the observed 2D surface brightness distributions of each galaxy and build host galaxy models that account for dust extinction. A description of our dynamical modeling formalism is described in Section \ref{sec:Dynamical Modeling}, and the results are presented in Section \ref{sec:Results}. We discuss the challenges and limitations of our measurements in Section \ref{sec:Discussion}, and conclude in Section \ref{sec: Conclusion}.

\section{Observations}
\label{sec:Observations}

\subsection{ALMA Observations}
\label{sec: ALMA Obs}

We obtained ALMA imaging of NGC 4786 and NGC 5193 from ALMA Programs 2015.1.00878.S and 2017.1.00301.S, respectively. NGC 4786 was observed on 23 July 2016 for approximately 21 minutes with a maximum baseline of 1110 m. The observation consisted of three spectral windows targeting continuum emission and one spectral window targeting the redshifted CO(2$-$1) emission line. The continuum windows had a channel resolution of 15.625 MHz and covered the following frequency ranges: $227.14-231.14$ GHz, $239.47-243.47$ GHz, and $241.78-245.78$ GHz. The emission line spectral window had a channel resolution of 3.906 MHz and spanned the frequencies between $225.14-228.89$ GHz. The $uv$-plane visibilities were reduced and calibrated in version 4.5.3 of the Common Astronomy Software Applications package (CASA; \citealp{2007ASPC..376..127M}), and then imaged into a data cube with 20 $\mathrm{km}\, \mathrm{s}^{-1}$ velocity channel spacings (with respect to the rest frequency of the CO(2$-$1) emission line at 230.538 GHz) using a robust parameter of 0.5. We chose a pixel size of $0\farcs{05}$ to adequately sample the synthesized beam's minor axis. The beam's position angle is $67.3^{\circ}$ measured East of North. The major axis full width at half maxmium (FWHM) of the synthesized beam is $0\farcs{35}$, whereas the minor axis has a FWHM of $0\farcs{27}$, giving it a geometric mean FWHM of $0\farcs{31}$. 

NGC 5193 was observed on 15 January 2018 for approximately 29 minutes with a maximum baseline of 2386 m. The observation targeted the redshifted CO(2$-$1) emission line along with a corresponding spectral window for the continuum. The emission line spectral window had a channel resolution of 3.904 MHz, and covered the frequency range of $226.84 - 228.71$ GHz. The continuum windows had a channel resolution of 31.25 MHz, and covered frequencies between $224.78 - 226.76$ GHz. The $uv$-plane visibilities were calibrated in CASA version 5.1.1, and then imaged into a data cube with 10 $\mathrm{km}\, \mathrm{s}^{-1}$ velocity channel spacings, with a pixel size of $0\farcs{035}$. The synthesized beam has a position angle of $63.1^{\circ}$ measured East of North, has a major axis FWHM of $0\farcs{33}$, and has a minor axis FWHM of $0\farcs{29}$, giving it a geometric mean FWHM of $0\farcs{31}$.

\begin{figure*}[ht]
    \centering
    \includegraphics[width=7in]{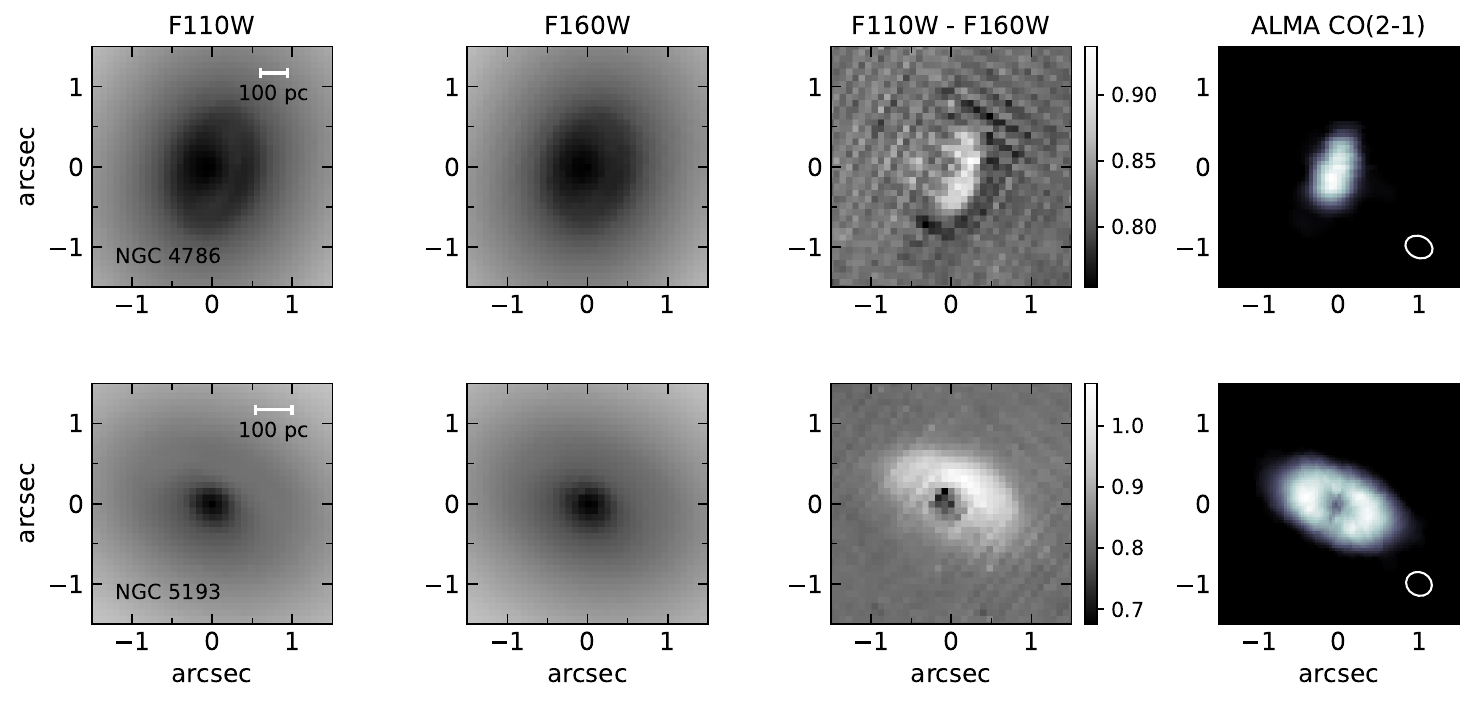}
    \caption{HST F110W ($J$-band), F160W ($H$-band), $\mathrm{F110W} - \mathrm{F160W}$ ($J-H$), and ALMA CO(2$-$1) images of NGC 4786 (top) and NGC 5193 (bottom) showing the co-spatial alignment of the gas and dust. The ALMA integrated flux density maps were created by summing channels in the data cubes that displayed visible CO emission. Pixels with emission were identified with an automatically generated mask by the \texttt{3DBarolo} program \citep{2015MNRAS.451.3021D}. In the $J-H$ maps, light regions correspond to redder colors and dark regions are bluer than the surrounding starlight. North is up and East is to the left in each image. Colorbars are shown in magnitude units.} 
    \label{fig:hst_alma_observations}
\end{figure*}

\subsubsection{Circumnuclear Disk Properties}
\label{sec:CircumnuclearDisksALMA}
A detailed description of the properties of these ALMA datasets was presented by \cite{PhD2018Boizelle}. We briefly describe some of their properties below. As seen in Figure \ref{fig:hst_alma_observations}, the CO emission is cospatial with the optical dust disk. The CO surface brightness extends about $0\farcs{65}$ in radius along the disk's major axis in NGC 4786 and about $1\farcs{2}$ for NGC 5193. Both disks display orderly rotation about their centers, with the projected LOS velocities reaching ${\sim}270 \,\mathrm{km}\,\mathrm{s^{-1}}$ in each disk. Examination of their respective position-velocity diagrams (PVDs) extracted along the major axis reveals that the CO velocities are relatively flat and decrease slightly towards their respective disk center. The PVDs also highlight a lack of CO-bright gas well within the expected BH SOIs. 

To incorporate the mass of the gas disk in the total gravitational potential of our dynamical models, we convert the integrated CO(2$-$1) flux measurements over each disk into $M_{\mathrm{gas}}$ profiles. These $M_{\mathrm{gas}}$ profiles are dominated by molecular hydrogen and helium and are calculated as $M_{\mathrm{gas}} = M_{\mathrm{H_2}}(1 + f_{\mathrm{He}})$, where we set $f_{\mathrm{He}} = 0.36$. 

The process of generating $M_{\mathrm{gas}}$ profiles starts with the construction of an integrated CO(2$-$1) flux map. To build this map, we multiply the data cube with a 3D mask generated by the \texttt{3DBarolo} program \citep{2015MNRAS.451.3021D} and sum the channel maps along the spectral axis to generate a 2D map of integrated flux density. Upon converting the map into units of integrated flux, we average the flux on elliptical annuli centered on the disk centers. If we sum the integrated flux across the entire region of each disk, we find $I_{\mathrm{CO(2-1)}} = (6.90 \pm 0.14) \, \mathrm{Jy} \, \mathrm{\kms}$ for NGC 4786 and $I_{\mathrm{CO(2-1)}} = (7.10 \pm 0.05) \, \mathrm{Jy} \, \mathrm{\kms}$ for NGC 5193. These statistical uncertainties are calculated through Monte Carlo simulations, but there is an additional 10\% systematic uncertainty that stems from the flux scale. For each elliptical annulus, the integrated CO(2$-$1) flux measurements are converted  into CO(1$-$0) luminosities using:
\begin{equation}
    L^{\prime}_{\mathrm{CO}} = 3.25 \times 10^7 S_{\mathrm{CO}}{\Delta v}\frac{D_L^2}{(1+z)^3\nu_{\mathrm{obs}}^2}\, \mathrm{K}\,\mathrm{\kms}\,\mathrm{pc}^2
\end{equation}
\citep{2013ARA&A..51..105C} assuming a CO(2$-$1)/CO(1$-$0) line ratio of 0.7 \citep{1999AJ....117.1995L}. Then, a mass of $\mathrm{H_2}$ is obtained by multiplying the CO(1$-$0) luminosities by $\alpha_{\mathrm{CO}} = 3.1\, M_{\odot} \,\, \mathrm{pc}^{-2}$ ($\mathrm{K}\, \mathrm{km}\,\mathrm{s}^{-1})^{-1}$ \citep{2013ApJ...777....5S} and then multiplying this result by 1.36 as described above to generate an estimate of $M_{\mathrm{gas}}$, though it should be noted that the most appropriate $\alpha_{\mathrm{CO}}$ value for ETGs is unknown, and thus the estimated $M_{\mathrm{gas}}$ values should be taken as approximations. We find $M_{\mathrm{gas}} = 6.9 \times 10^7 \, M_{\odot}$ for the disk in NGC 4786 and $M_{\mathrm{gas}} = 3.9  \times 10^7 \, M_{\odot}$ in NGC 5193. Along with the unknown ideal value of $\alpha_{\mathrm{CO}}$, the uncertainty in $D_L$ contributes ${\approx} 15\%$ based on the range of redshift-independent distances in NED for NGC 4786 and ${\approx}7\%$ from the SBF distance measurement to NGC 5193 by \cite{2003ApJ...583..712J}. This distance uncertainty is in addition to the $10\%$ systematic and $(1-2)\%$ statistical uncertainty associated with the integrated flux measurements. As will be discussed in  Section \ref{sec:ErrorAnalysis}, the inclusion or exclusion of the gas mass in the total gravitational potential of the system only contributes a small amount to the total error budget on $\mbh$ in each galaxy.

\subsection{HST Observations}
\label{sec:HSTdata}

All the HST data used in this paper can be found in MAST: \dataset[10.17909/gf2w-xv03]{http://dx.doi.org/10.17909/gf2w-xv03}. For each galaxy, we retrieved F110W ($J$-band) and F160W ($H$-band) images taken with the Wide Field Camera 3 (WFC3). For NGC 4786, the $H$-band image was taken with a four-point dither pattern with four separate exposures that lasted 249 seconds each. The $H$-band images for NGC 5193 employed a similar observation strategy with four separate exposures lasting 299 seconds each. For the $J$-band images, a two-point dither pattern was used and two separate 249-second exposures were taken for NGC 4786, whereas two 128-second exposures were taken for NGC 5193. Further details regarding the data mosaicing process, and construction of dust-masked MGEs for a larger sample of ETGs will soon be available (Davidson et al., in prep), but we summarize the key steps below. 

We processed our data through the \texttt{calwf3} pipeline and used \texttt{AstroDrizzle} to combine and align the separate exposures. To start, we drizzled and aligned the flat-fielded $H$-band images to a pixel scale of $0\farcs{08}$, and used this image as the reference image when drizzling and aligning the $J$-band image. We determined offsets between the $H$ and $J$-band images from the luminosity-weighted galaxy center coordinates of each image, and interpolated to align them to within ${\sim}0.2$ subpixels of accuracy based on inspection of the $J-H$ maps. Additionally, we constructed \texttt{TinyTim} \citep{krist04} model point-spread functions (PSFs) that were dithered and drizzled in the same fashion as the $H$-band image. These PSFs are needed to construct the host galaxy models.

\section{Host Galaxy Modeling}
\label{sec:HostGalaxyModeling}

\begin{figure*}[ht]
\centering
\includegraphics[width=7.0in]{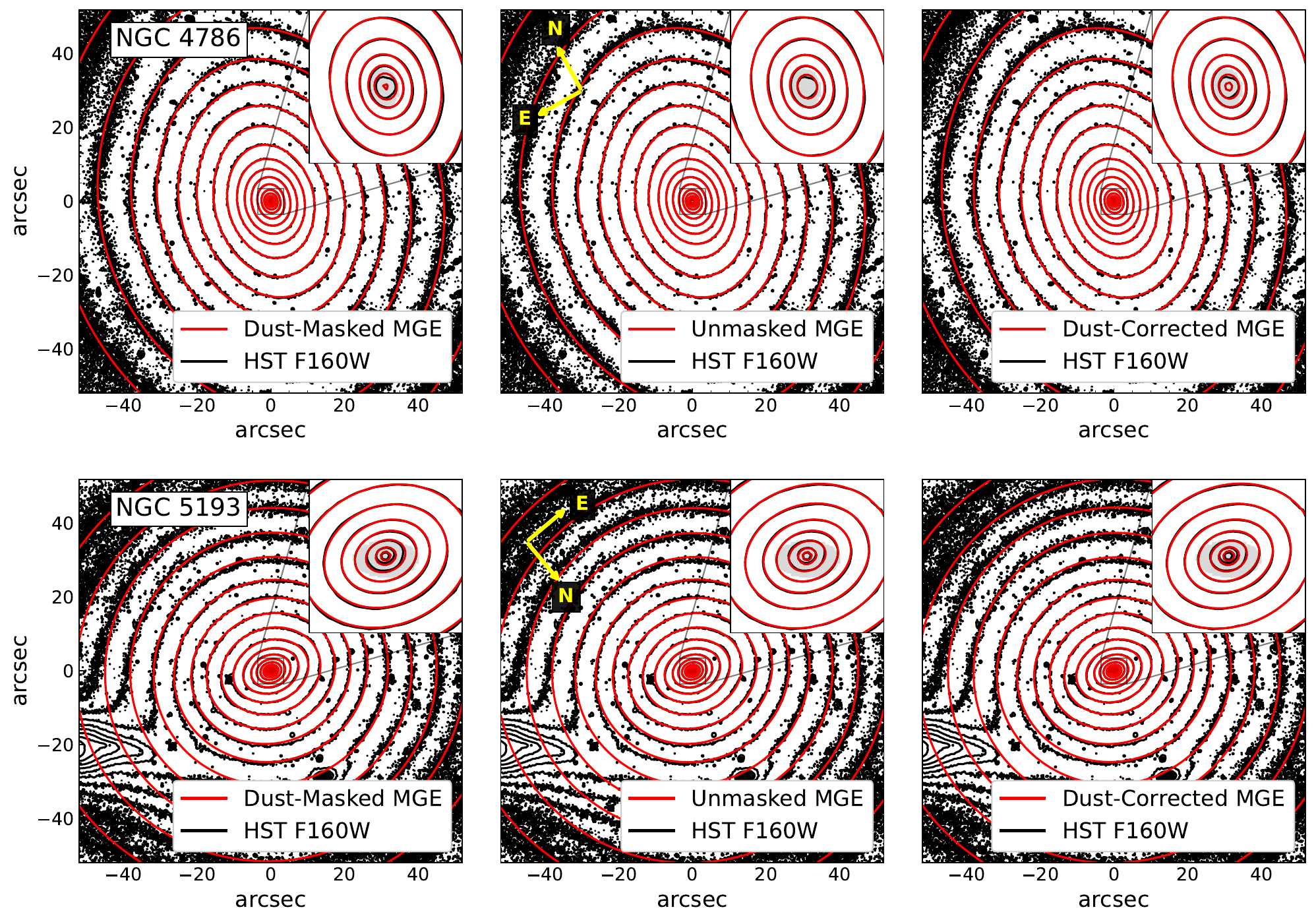}
\caption{2D isophote maps comparing the observed HST WFC3 F160W isophotes to those of our three MGEs for NGC 4786 (top) and NGC 5193 (bottom). Black contours represent isophotes from the F160W images, while red contours are for the MGE models. For each image, the central ${\approx}100\arcsec \times 100 \arcsec{}$ region is displayed with an inset of the innermost $3\farcs{5} \times 3\farcs{5}$ region in the top right corner. The gray ellipse shown within each inset indicates the size and orientation of the circumnuclear dust disk. Arrows in the middle panels indicate the orientation of North and East for each galaxy.}
\label{fig:Isophotes}
\end{figure*}

\begin{figure*}[t]
\centering
\includegraphics[width=3.2in]{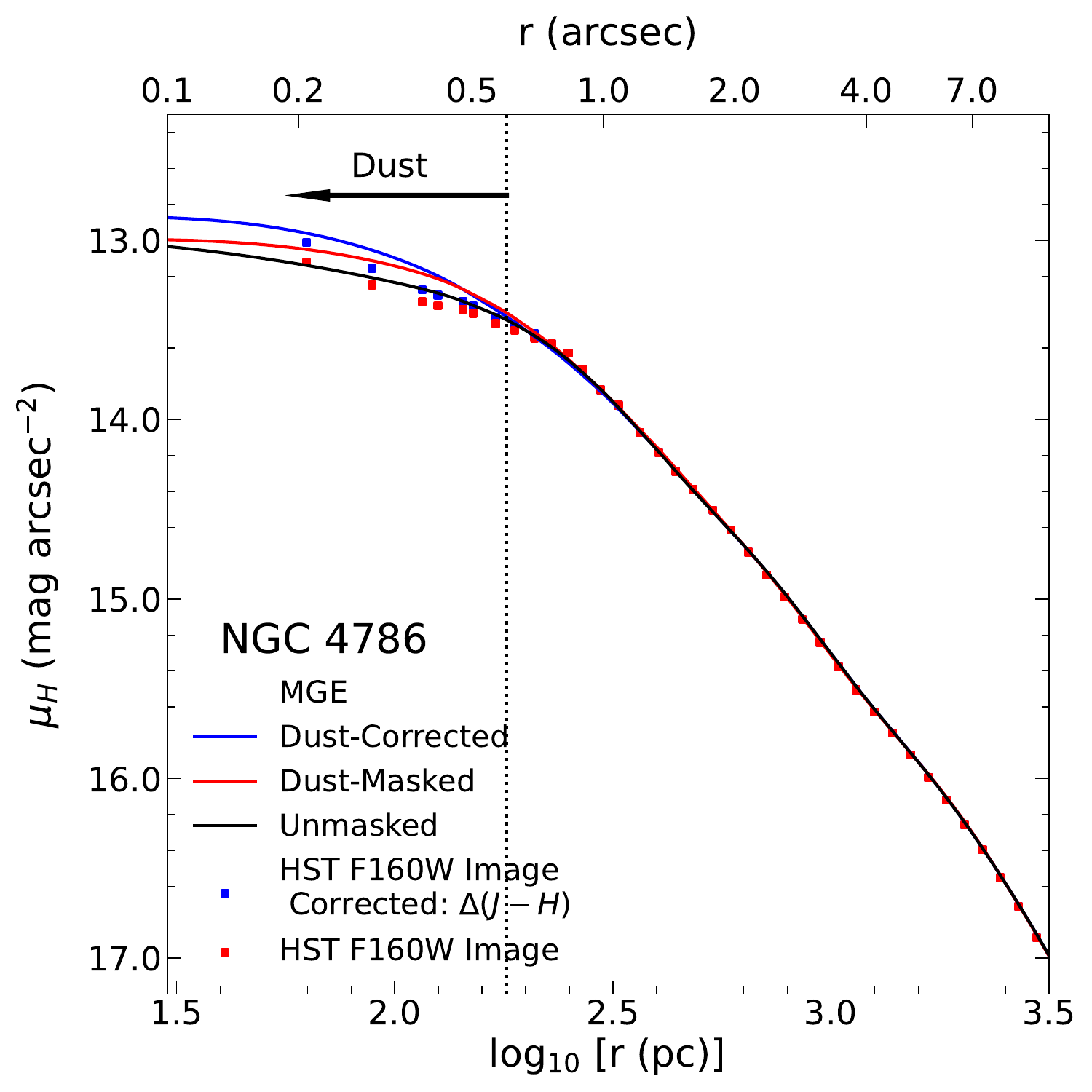}
\includegraphics[width=3.2in]{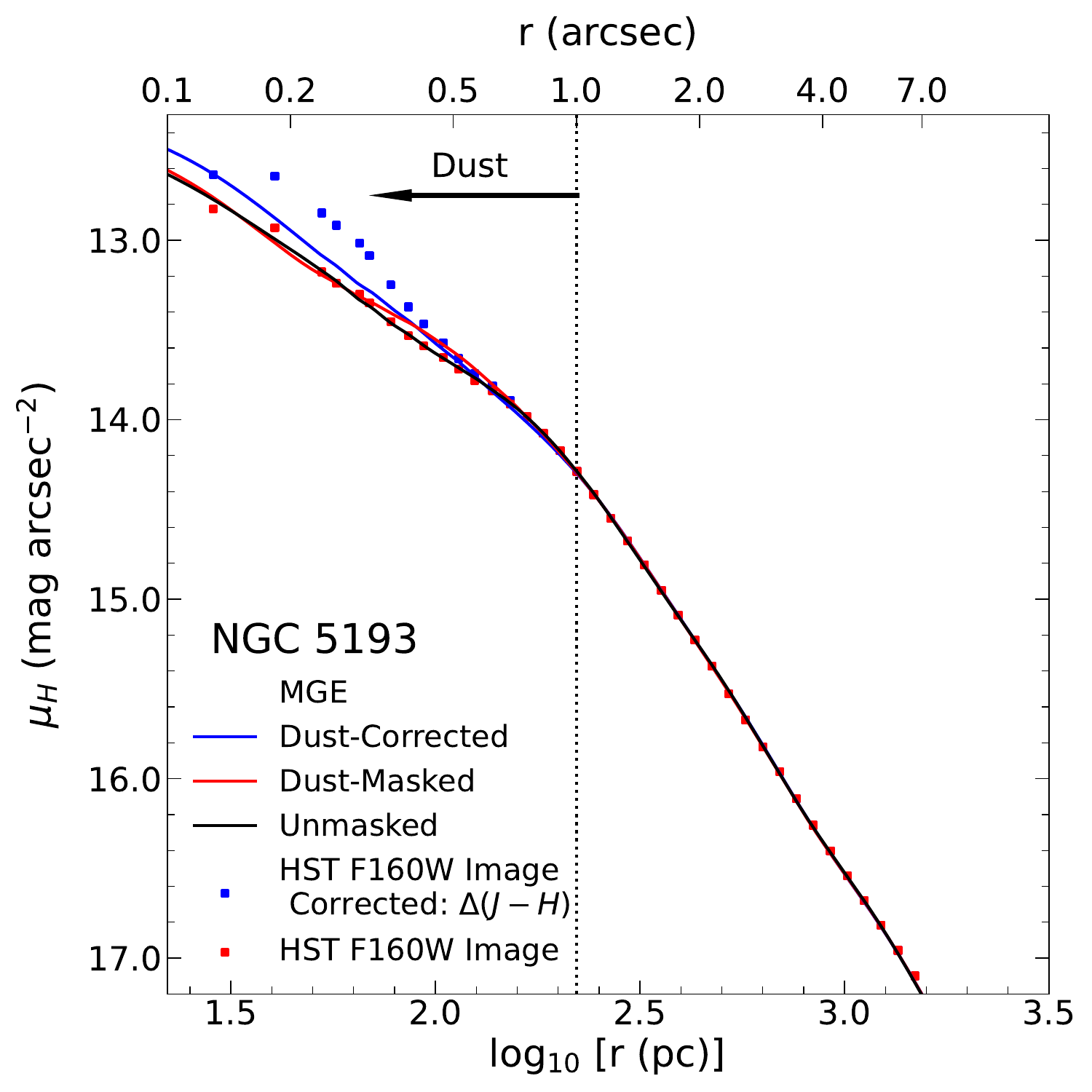}
\includegraphics[width=3.2in]{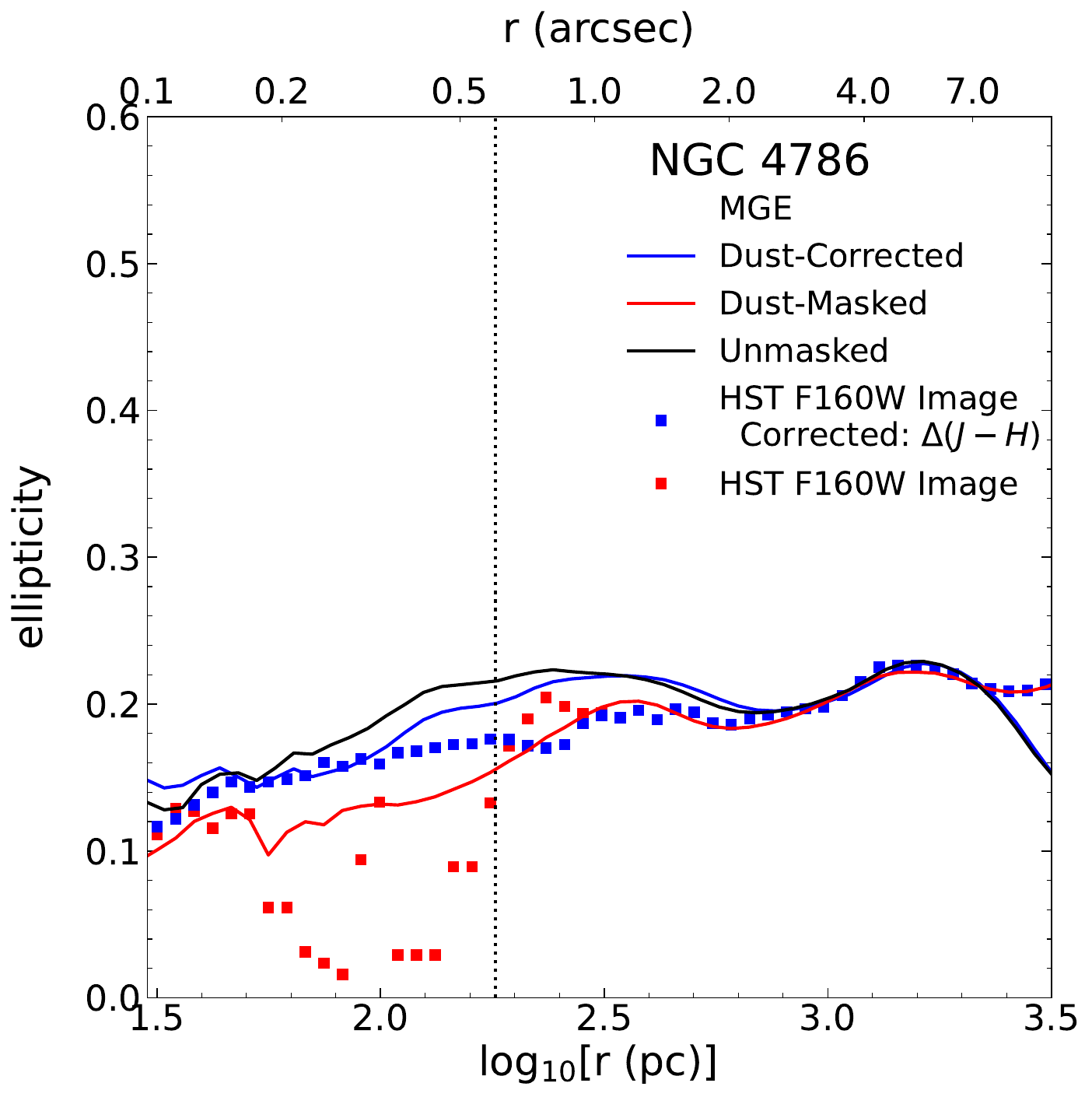}
\includegraphics[width=3.2in]{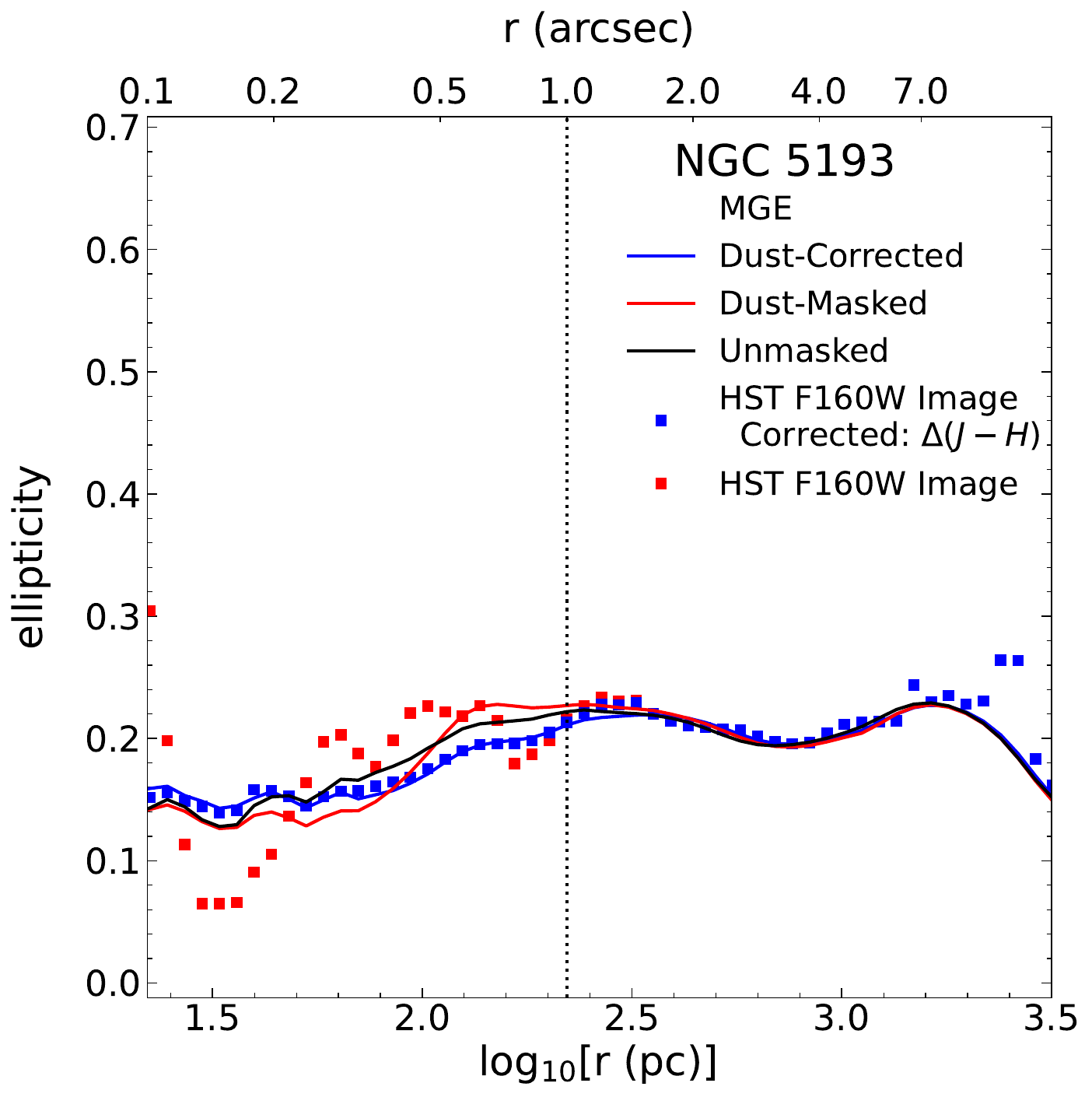}
\caption{(\textit{Top}) A comparison of the observed and modeled $H$-band surface brightness profiles of NGC 4786 (left) and NGC 5193 (right). The surface brightness measurements are made with the Python-implementation of the \texttt{sectors\_photometry} routine \citep{2002MNRAS.333..400C} which performs photometry along evenly spaced sectors from the major axis to the minor axis and averages measurements over the four quadrants of the image.
\textit{(Bottom)} A comparison of the radial ellipticity of the observed $H$-band photometry and the 2D MGE models for NGC 4786 (left) and NGC 5193 (right). We use the \texttt{photutils.isophote} routine \citep{1987MNRAS.226..747J,larry_bradley_2023_7946442} which fits elliptical isophotes to a galaxy image to determine the ellipticity of each isophote along the semi-major axis.
For each panel, the red squares are the observed values from the $H$-band image, while blue squares are dust-corrected values described in Sections \ref{sec:DustCorrectedMGE}. The solid lines in each panel correspond to profiles extracted along the major axis for each of our 2D MGE models. 
Red lines are for dust-masked MGE models whereas black and blue lines represent dust-unmasked and dust-corrected MGEs, respectively. The dashed lines indicate the dust disk edge and the arrows indicate that the dust extends down to the nucleus.}
\label{fig:SurfaceBrightnessProfiles}
\end{figure*}

An accurate host galaxy model that accounts for the mass of the galaxy's stars is crucial when measuring the mass of a central BH. These host galaxy models are constructed by modeling and deprojecting the observed 2D surface brightness profiles from the drizzled $H$-band images. Given the presence of a circumnuclear dust disk, we chose to model the galaxy's surface brightness in the $H$-band because dust attenuation is reduced at longer wavelengths. We used the Multi-Gaussian Expansion (MGE) method \citep{1994A&A...285..723E,2002MNRAS.333..400C} to parameterize the surface brightness of both galaxies with a sum of concentric Gaussian functions. 

\subsection{Dust-Masked and Unmasked MGE Models}
\label{sec:DustMaskedUnmaskedMGEs}
We built three unique MGE models for NGC 4786 and NGC 5193 following an approach similar to that used by \cite{2021ApJ...919...77C}. The three models correspond to an unmasked, dust-masked, and dust-corrected MGE model that account for the effects of dust differently, and are used to assess the systematic impact of the chosen MGE model on our BH mass measurement. 

As seen in Figure \ref{fig:Isophotes}, the $H$-band dust attenuation does not appear severe in either NGC 4786 and NGC 5193, hence we first explored both unmasked and dust-masked MGE models to explore the impact on the measurement of $\mbh$. Prior to fitting MGEs to the drizzled $H$-band image of each galaxy, we isolated the host galaxy light in each image from extraneous sources such as neighboring galaxies, foreground stars, cosmic rays, and detector artifacts by masking these objects. In addition, we constructed $J-H$ maps for each galaxy in order to better identify and mask out regions where dust obscuration was highest, typically corresponding to areas where $J-H > 0.88$ mag, or equivalently where the color excess is $0.08$ mag higher than in regions just outside the disk. For all reported $H$ and $J$-band magnitudes in this work, we use the Vega magnitude system.

We first modeled the 2D surface brightness with routines from the \texttt{MgeFit} package in Python \citep{2002MNRAS.333..400C}. The components of this initial MGE were then used as initial guesses for a second MGE fit using the \texttt{GALFIT} program \citep{2002AJ....124..266P}. We chose to use \texttt{GALFIT} for our final MGEs because it allows for an asymmetric 2D PSF to be incorporated in the modeling process, in contrast to \texttt{MgeFit} which requires decomposing the PSF into a sum of circular Gaussian functions when used in the MGE construction. For both programs, we accounted for the blurring due to the $H$-band PSF by incorporating the \texttt{TinyTim} $H$-band PSF models we built. In addition, our MGEs account for a foreground $H$-band Galactic reddening of $A_H = 0.019$ mag for NGC 4786 and $A_H = 0.029$ mag for NGC 5193 \citep{2011ApJ...737..103S}. Our MGE solutions contain fourteen Gaussian components for NGC 4786 and eight Gaussian components for NGC 5193 and are listed in Tables \ref{tab:MGE_NGC4786_Unmasked_DustMasked} and \ref{tab:MGE_NGC5193_Unmasked_DustMasked} in the Appendix. 

We carefully considered whether to include a PSF component in \texttt{GALFIT} to account for a potential unresolved nuclear source of non-stellar origin. Optical spectra of the nuclear regions showed no evidence of prominent emission lines typically associated with an active galactic nucleus in either galaxy \citep{2009MNRAS.399..683J}. While the NGC 5193 $H$-band surface brightness profile exhibits a cuspy nature, an examination of the galaxy center in multiple wavelength filters revealed that the central light is radially extended and not point-like. Based on these findings, we decided not to include a central PSF component in our model for either galaxy's central surface brightness distribution. 

Our MGE models assume that the galaxy has an oblate and axisymmetric shape, and that each Gaussian component shares the same center and position angle. While individual Gaussian components may not correspond to physically distinct galaxy components, their projected axial ratios $q_k^{\prime}$ may converge on values below $\cos(i)$ in the MGE optimization process, where $i$ is the inclination of the cirumnuclear disk. A useful proxy for the inclination is $i = \arccos(b/a)$ where $b/a$ is the observed axial ratio of the disk as measured from the HST images. This proxy typically agrees with kinematic inclinations derived from dynamical models to within ${\sim}5^{\circ}$ based on previous studies \citep{2019ApJ...881...10B,2021ApJ...908...19B,2021ApJ...919...77C,2022ApJ...934..162K}, and so we set a lower bound on the possible MGE component axial ratios of 0.69 and 0.75 for NGC 4786 and NGC 5193. This enables deprojection of the MGEs down to inclinations as low as $46^{\circ}$ and $51^{\circ}$, respectively.

Examination of the 2D isophotes in Figure \ref{fig:Isophotes} shows that the model isophotes are an excellent match to those seen in the $H$-band data out to ${\sim}100 \arcsec{}$. Within the central dusty regions, the observed $H$-band isophotes for both NGC 4786 and NGC 5193 remain relatively symmetrical and are modeled well by their dust-masked MGEs. Extracting major axis surface brightness profiles from both the MGE model and the $H$-band data in Figure \ref{fig:SurfaceBrightnessProfiles} also shows good agreement at intermediate radii, though discrepancies within the dusty region are noticeable. Despite slight mismatches in the model and data surface brightness profiles, the dust in NGC 4786 and NGC 5193 appears to have a less noticeable impact on the observed $H$-band surface brightness distribution in each galaxy in comparison to what has been seen in previous work, such as in the ETGs NGC 1380 and NGC 6861, where the $H$-band isophotes become non-elliptical and asymmetric within the dust disk  \citep{2022ApJ...934..162K}.
Additionally, we show the ellipticity as a function of semi-major axis for the $H$-band data and the various MGE models in the bottom panels of Figure \ref{fig:SurfaceBrightnessProfiles}.
The ellipticity profiles of the unmasked and dust-corrected MGE models are in good agreement with their respective images.
The unmasked $H$-band photometry is poorly behaved within the dust lane.
The ellipticity profiles of the dust-corrected images are better behaved than those of the unmasked $H$-band images.

\subsection{Dust-Corrected MGE Models}
\label{sec:DustCorrectedMGE}
The final MGE we created for each galaxy involved fitting an MGE model to a dust-corrected $H$-band image. The process of developing this MGE model follows methods described by \cite{2019ApJ...881...10B,2021ApJ...908...19B}, \cite{2021ApJ...919...77C}, and \cite{2022ApJ...934..162K}. We summarize the key steps below. 

We fit a 2D Nuker model \citep{1997AJ....114.1771F} to the innermost $10\arcsec{} \times 10\arcsec{}$ of the drizzled $H$-band image using \texttt{GALFIT}. This fit included the mask we used for the dust-masked MGE, and acts as the starting point of our dust correction. We use a 2D Nuker model to fit the central region of the galaxy as we can tune its parameters to create dust-corrected HST images corresponding to specific values of $A_H$ at the center. We prefer using Nuker models to perform these adjustments as opposed to simply adjusting the observed flux values in the HST image and fitting an MGE to this adjusted image directly because the Nuker models not only provide a way to handle the adjustments at the center, but also provide an estimate of the intrinsic surface brightness in dust-affected regions. This is in contrast to simply adjusting the observed surface brightness values at the galaxy center in the HST image and fitting an MGE to this new image, as the adjustment can create sharp or discontinuous features in the surface brightness profile that are unphysical and can manifest in the resultant MGE model.

The Nuker model fits in \texttt{GALFIT} include the $H$-band PSFs, so the resulting solutions correspond to intrinsic parameters. Nuker models have been shown to accurately model the surface brightness distribution within the innermost few arcseconds of early-type galactic nuclei, and they characterize this distribution with an inner and outer power-law profile \citep{2007ApJ...664..226L}. Mathematically, the Nuker law has the following form: ${I(r) = I_b2^{\frac{\beta-\gamma}{\alpha}}(r/r_{\mathrm{b}})^{-\gamma}[1 + (r/r_{\mathrm{b}})^{\alpha}]^{\frac{\gamma - \beta}{\alpha}}}$, with $\gamma$ and $\beta$ representing the slopes of inner and outer power laws, respectively. The transition between these two regimes occurs at a given break radius, $r_{\mathrm{b}}$, and the sharpness of this transition is described by the parameter $\alpha$. For NGC 4786, our \texttt{GALFIT} optimization converged on the following values: an ellipticity of $\epsilon = 0.20$, $\alpha = 1.94,\,\beta = 1.56$,\,$\gamma=0.00$, and $r_{\mathrm{b}}=0\farcs{46}$, which are typical values for cored-elliptical galaxies \citep{2007ApJ...664..226L}. These cores are hypothesized to originate through scouring by massive supermassive BH binaries \citep{2002ApJ...566..801R,2014ApJ...782...39T}. The values for our NGC 5193 Nuker model were: an ellipticity of $\epsilon = 0.22$, $\alpha = 6.28,\,\beta = 1.40$,\,$\gamma=0.70$, and $r_{\mathrm{b}}=1\farcs{39}$, which are values characteristic of power law galaxies \citep{2007ApJ...664..226L}.

The next step involves estimating how much extinction to the observed $H$-band stellar light there is at the center of the galaxy. We follow the approach described by \cite{2022ApJ...934..162K}, which uses the observed $J-H$ color map of each galaxy to determine an estimate of $A_H$, the extinction of the $H$-band stellar light originating behind the disk.  First, we determined a median $J-H$ color of 0.81 mag and 0.82 mag outside the dust disks of NGC 4786 and NGC 5193, respectively, and we determined the color excess, $\Delta(J-H) = (J-H) - (J-H)_{\mathrm{median}}$ as a function of position along the disk's major axis, averaging over a width of 4 pixels. We note that the $J-H$ map of NGC 5193 is indicative of a central hole in the dust distribution and is supported by the CO(2$-$1) moment 0 map which displays a deficit of central CO emission as well. 

To establish a relationship between extinction and color excess, we used Equations (1) and (2) from \cite{2019ApJ...881...10B} to generate a curve of $\Delta(J-H)$ as a function of $V$-band extinction, $A_V$ (see Figure 4 of \citealp{2022ApJ...934..162K}). This assumes the \cite{2017MNRAS.472.1286V} embedded-screen model, which effectively models the circumnuclear dust disk as a thin, inclined disk that bisects the galaxy. Along a given LOS, the fraction of light that originates in front of the disk ($f$) is unaffected by dust, while the fraction behind it $(b)$ is obscured by screen extinction. The ratio of observed to intrinsic integrated $H$-band stellar light is represented mathematically as $F_{\mathrm{observed}}/F_{\mathrm{intrinsic}} = f\, + \,b[10^{-A_H/2.5}]$. This is from Equation (1) in \cite{2019ApJ...881...10B}, and assumes an intrinsically thin disk, where fractional disk thickness $w = 0$. Along the major axis of each disk, the fractions of light originating in front of and behind of the dust disk are assumed to be equal ($f = b = 0.50$). 

The next key step in this process is converting the observed $\Delta(J-H)$ as a function of position along the disk's major axis into values of $A_H$. Using our curve of $\Delta(J-H)$ versus $A_V$, we can associate a unique value of $A_V$ (as well as $A_H$) to an observed $\Delta(J-H)$ value. As seen in Figure 4 of \cite{2022ApJ...934..162K}, this is only valid up to a given turnover point. This is due to the fact that at large ($A_V > 5$ mag) optical depths, variations in color begin to rapidly diminish, and so the same value of $\Delta(J-H)$ can correspond to both low and high $A_V$ values. Following the procedure outlined by \cite{2022ApJ...934..162K}, we assumed the lower $A_V$ value, as the higher value implies that effectively all of the light originating behind the disk is lost due to extinction. We fit the color excess curve with a third-order polynomial up to the turnover point. To generate predictions of $A_V$ as a function of $\Delta(J-H)$, we use this polynomial's inverse. Then, we found the lower $A_V$ values corresponding to the observed $\Delta(J-H)$ along the disk's major axis. Finally, we set $A_H = 0.175A_V$ based on the standard interstellar extinction law described in \cite{1985ApJ...288..618R}, which gives us a unique $A_H$ value for each observed color excess along the major axis. As stated earlier, this $A_H$ value applies only to light originating behind the disk.

Our simple dust correction implies $A_H = 0.22$ mag and $A_H = 0.18$ mag at the centers of the circumnculear disks in NGC 4786 and NGC 5193, which correspond to a reduction of approximately 20\% and 15\% of the stellar light originating behind the disk. We note that a proper treatment of determining the intrinsic stellar light distribution in the presence of a dusty circumnuclear disk likely requires the usage of radiative transfer models \citep{2013A&A...550A..74D,2015A&C.....9...20C} that account for the combined effects of extinction, light scattering, and the geometry of the dust disk itself. Even still, our simple method gives us a relatively straightforward way of producing an estimate of the assumed extinction, and consequently, the impact it has on the measured value of $\mbh$.

With these estimated values of $A_H$ for each galaxy, we proceeded to mask the entirety of the dust disk in the $H$-band images except for the central nine pixels. This was done to anchor the model fit to the observed values at the center. The fluxes of these nine pixels are subsequently boosted by a factor of ${(0.50\,+ \,0.50
\times 10^{-A_H/2.5})^{-1}}$. We also tested this procedure with the central four pixels as well and found no significant difference between the two cases. With the entire dust disk masked out except for the pixels that have had their flux values increased, we re-fit the central $10\arcsec{} \times 10\arcsec{}$ region again with a new Nuker model, but fixed the values of $\alpha, \beta$, and $r_{\mathrm{b}}$ to their values from the previous Nuker model to retain the larger scale properties outside of the dusty region. We then adjusted the inner slope parameter $\gamma$ to find an optimal value where the central pixels of the Nuker model are nearly equal to the scaled flux values of the $H$-band image. For NGC 4786, this value is $\gamma = 0.11$, and for NGC 5193 it is $\gamma = 0.75$. 

With this new Nuker model, the final steps in our dust correction process are to replace the pixels within the dust disk region in the $H$-band image with the corresponding pixels in the Nuker model, and to fit this dust-corrected $H$-band image with a new MGE. As will be discussed in Section \ref{sec:Results}, these correspond to the MGEs that are used in our fiducial dynamical models. The dust-corrected MGE components are displayed in Table \ref{tab:MGE_fiducialcomponents}.

\begin{deluxetable*}{ccccccc}[ht]
\tabletypesize{\small}
\tablecaption{$H$-band Dust-Corrected MGE Parameters}
\tablewidth{0pt}
\tablehead{
\multicolumn{1}{c}{$k$} &
\colhead{$\log_{10}$ $I_{H, k}$ ($L_{\odot}\, \mathrm{pc}^{-2}$)} &
\colhead{$\sigma_{k}^{\prime}$ (arcsec)} &
\multicolumn{1}{c}{$q_{k}^\prime{}$} & \colhead{$\log_{10}$ $I_{H, k}$ ($L_{\odot}\, \mathrm{pc}^{-2}$)} &
\colhead{$\sigma_{k}^{\prime}$ (arcsec)} & \multicolumn{1}{c}{$q_{k}^\prime{}$}\\[-1.5ex]
\multicolumn{1}{c}{(1)} & \colhead{(2)} & \colhead{(3)} & 
\multicolumn{1}{c}{(4)} & \colhead{(5)} & \colhead{(6)} & 
\multicolumn{1}{c}{(7)}}
\startdata
  & \multicolumn{3}{c}{\bf NGC 4786} & \multicolumn{3}{c}{\bf NGC 5193}\\ \hline 
1 & 4.472  & 0.226 & 0.830  & 5.274 & 0.094  & 0.768  \\
2 & 4.474 & 0.545 & 0.784  & 4.550  & 0.334  & 0.794 \\
3 & 4.152 &  1.282  & 0.820  & 4.338  & 0.933 & 0.756 \\
4 & 3.580  &  2.654 & 0.756  & 3.982 & 2.015  & 0.814  \\
5 & 3.451 & 4.758 & 0.794 & 3.436 & 4.716  & 0.750  \\
6 & 2.638 & 5.479  & 0.949 & 3.020 & 10.017  & 0.848   \\
7 & 2.570 & 8.493  & 0.690 & 2.472 & 19.064 & 0.984  \\
8 & 2.716 & 12.896 & 0.690 & 1.654  & 45.548  & 0.962   \\
9 &  2.115 & 16.166  & 0.999  & $\cdots$  & $\cdots$ & $\cdots$   \\
10 & 2.217 & 21.964  & 0.690  & $\cdots$ & $\cdots$ & $\cdots$  \\
11 & 1.558  & 29.934 & 0.975  & $\cdots$ & $\cdots$  & $\cdots$  \\
12 & 1.454  & 56.294  & 0.690  & $\cdots$ & $\cdots$  & $\cdots$  \\
13 & 1.085  & 32.662  & 0.690  & $\cdots$ & $\cdots$  & $\cdots$  \\
14 & 1.062  & 125.555  & 0.958  & $\cdots$ & $\cdots$  & $\cdots$  \\
\enddata
\tablecomments{NGC 4786 and NGC 5193 dust-corrected MGE solutions created from the combination of HST $H$-band images and best-fitting \texttt{GALFIT} Nuker models. As described in Section \ref{sec:Results}, these two MGEs are used in the dynamical models with the lowest $\chi^2$. The first column is the component number, the second is the central surface brightness corrected for Galactic extinction and assuming an absolute solar magnitude of $M_{{\odot},H} = 3.37$ mag \citep{2018ApJS..236...47W}, the third is the Gaussian standard deviation along the major axis, and the fourth is the axial ratio. Primes indicate projected quantities.
The \texttt{GALFIT} MGE position angle found for NGC 4786 was $-17.0^\circ$ East of North and for NGC 5193 was $71.2^\circ$ East of North.}
\label{tab:MGE_fiducialcomponents}
\end{deluxetable*}

\section{Dynamical Modeling}
\label{sec:Dynamical Modeling}
The BH masses are measured by modeling the observed gas kinematics in the ALMA data cubes with a thin, rotating disk model. Our models use a minimum of nine free parameters which include BH mass $M_{\mathrm{BH}}$, $H$-band mass-to-light ratio $\Upsilon_H$, disk inclination angle $i$, disk position angle $\Gamma$, disk dynamical center $(x_{\mathrm{c}},y_{\mathrm{c}})$, CO flux normalization factor $F_0$, and turbulent velocity dispersion $\sigma_0$. The details of our modeling process are described by \cite{2022ApJ...934..162K}, but we summarize the key aspects here. 

We create synthetic data cubes that model the observed CO line profiles and fit them directly to the ALMA data cubes. To start, we build a model circular velocity field on a grid that is oversampled by a factor of $s = 3$ relative to the size of the spatial pixels in the data cube. The circular velocity at each grid point is a function of radius and is determined by the quadrature sum of the velocity contributions from the BH (treated as a point source with mass $\mbh$), the host galaxy's stars, and from the gas disk itself. We neglect the contribution of dark matter as the enclosed mass profiles on scales comparable to the circumnuclear disks are thought to be dominated by the stars. To derive the circular velocity profile due to the stellar mass, we deprojected the MGEs of each galaxy described in Section \ref{sec:HostGalaxyModeling} using routines from the \texttt{JamPy} package \citep{2008MNRAS.390...71C} under the assumptions that NGC 4786 and NGC 5193 are oblate, axisymmetric, and have inclination angles of $70^{\circ}$ and $60^{\circ}$, respectively, based on initial gas-dynamical modeling fits.
Each of these deprojections represents a possible three-dimensional luminous density for the galaxy.
With each model iteration, the stellar mass circular velocity profiles are scaled by $\sqrt{\Upsilon_H}$, the square root of the $H$-band mass-to-light ratio. 

The circular velocity contribution of the gas disk is obtained by numerically integrating projected gas mass surface densities that are calculated on the same annular regions described in the calculation of $M_{\mathrm{gas}}$ in Section \ref{sec:CircumnuclearDisksALMA}. We assumed the gas was distributed in a thin disk and determined the midplane circular velocity of a test particle orbiting in the disk's gravitational potential using Equation 2.157 in \cite{2008gady.book.....B}. 

For a given disk inclination angle $i$, position angle $\Gamma$, and assumed (fixed) distance to the galaxy, $D$, we calculate the LOS projection of the model velocities as seen on the plane of the sky. At each point in the disk, we model the emergent CO line profile as a Gaussian and use the projected LOS velocity and an assumed spatially uniform turbulent velocity dispersion term, $\sigma_{0}$, to build the model line profile. Given that the spectral axis of the ALMA data cubes is in frequency units, we transform the projected LOS velocity and velocity dispersion into a frequency line centroid and line width using the redshift, $z_{\mathrm{obs}}$ (related to the free parameter of the systemic velocity of the galaxy through $v_{\mathrm{sys}}/c = z_{\mathrm{obs}}$). 

Once the model data cubes are constructed, the Gaussian line profiles must be weighted by a CO flux map and each image slice needs to be convolved with the ALMA synthesized beam. The flux map is created by using a 3D mask generated by the \texttt{3DBarolo} program \citep{2015MNRAS.451.3021D}, multiplying this mask with the ALMA data cube, and summing this product along the spectral axis to form a 2D image. To deconvolve this image, we use an elliptical Gaussian PSF with major and minor FWHMs that match the specifications of the ALMA synthesized beam, and applied five iterations of the Richardson-Lucy algorithm implemented in the \texttt{scikit-image} Python package \citep{van2014scikit}. To weight the line profiles on the oversampled grid scale, we divide the CO flux map so that a pixel in the map is divided into $s^2 = 9$ subpixels and is normalized so that the subpixels corresponding to the same original pixel have equal fluxes. The model is then block-averaged down to the original ALMA scale and has each of its slices convolved with the model synthesized beam using the \texttt{convolution} implementation in the \texttt{astropy} framework \citep{astropy:2013,astropy:2018}.

The final step in the modeling process is to minimize $\chi^2$ between our model and the ALMA data. We compute $\chi^2$ on elliptical spatial regions centered on the disk centers shown in Figures \ref{fig:NGC4786Moments} and \ref{fig:NGC5193Moments}. Given the small number of resolution elements across each disk, we opted to fit our models to nearly the full extent of each disk, though we explore the systematic impact of this choice in Section \ref{sec:ErrorAnalysis}. Our elliptical fitting regions have an axial ratio of $q = 0.50$ and semimajor axis lengths of $a = 0\farcs{55}$ and $a=1\farcs{0}$ for NGC 4786 and NGC 5193, respectively. Additionally, to mitigate the impact of neighboring pixel-to-pixel correlation, we further block-averaged the cubes by a factor of 4. On these block-averaged scales, we minimized $\chi^2 = \sum_{i = 1}^{N} \left[(d_i - m_i)/\sigma_i\right]^2$, where $d_i, m_i$, and $\sigma_i$ represent the flux density of the 3D data, model, and noise cubes at a given pixel. The $\sigma_i$ values are determined from a 3D noise model that was generated for each galaxy described in detail in Section 4.2.1 of \cite{2022ApJ...934..162K}, but we briefly describe its construction here. We first calculated the noise in each frequency channel as the standard deviation of emission-free pixels on the block-averaged scale. This is done in a version of the data cube prior to primary beam correction as the noise is spatially uniform at this stage. Then, we took the primary beam cube generated during data calibration and reduction and block-averaged it by the same factor. The 3D noise model is then generated by dividing the uniform rms noise value by the primary beam for each frequency channel.

We fit our dynamical models to 31 frequency channels in the NGC 4786 data cube that correspond to velocities of $|v_{\mathrm{LOS}} - v_{\mathrm{sys}}| \leq {\sim}300 \,\mathrm{\kms}$. This velocity range slightly extends past the channels with visible CO emission.  After block-averaging our model and data cubes, the elliptical spatial region contains 12 pixels per channel, or a total of 372 data points in the entire model fit. For the NGC 5193 data cube, we fit our models to 55 spectral channels corresponding to $|v_{\mathrm{LOS}} - v_{\mathrm{sys}}| \leq {\sim}280 \,\mathrm{\kms}$. On the block-averaged scale, the elliptical spatial region contains 58 data points per channel which gives an overall total of 3190 data points used in the $\chi^2$ minimization.

\section{Results}
\label{sec:Results}

\begin{deluxetable*}{ccccccccccc}[t]
\tabletypesize{\small}
\tablecaption{Dynamical Modeling Results}
\tablewidth{0pt}
\tablehead{
\colhead{Model} & 
\colhead{MGE} &
\colhead{\mbh} & 
\colhead{$\Upsilon_H$} & 
\colhead{$i$} &
\colhead{$\Gamma$} &
\colhead{$\sigma_0$} &
\colhead{$v_{\mathrm{sys}}$} &
\colhead{$F_0$} & 
\colhead{$\chi^2_{\nu}$}
\\[-1.5ex]
\colhead{} & 
\colhead{} &
\colhead{($10^8 \,M_{\odot}$)} & 
\colhead{($M_{\odot}/L_{\odot}$)} & 
\colhead{($^{\circ}$)} &
\colhead{($^{\circ}$)} &
\colhead{$(\mathrm{km}\,\mathrm{s^{-1}})$} &
\colhead{$(\mathrm{km}\,\mathrm{s^{-1}})$} &
\colhead{}
} 
\startdata
\hline \multicolumn{9}{c}{\textbf{NGC 4786}} \\
A &  Unmasked & 3.9 & 2.76 & 70.8 & 162.5 & 10.8 & 4620.47 & 1.56 & 1.488\\
B & Dust-Masked & 5.8 & 1.98 & 70.8 & 164.4 & 9.3 & 4621.63 & 1.56 & 1.449 \\
C & Dust-Corrected & 5.0 & 1.80 & 69.3 & 166.6 & 9.9 & 4621.26 & 1.54 & 1.421 \\
& & (0.2) & (0.04) & (0.7) & (1.2) & (3.0) & (1.10) & (0.03) & \\
\hline \multicolumn{9}{c}{\textbf{NGC 5193}} \\
D & Unmasked & 1.5 & 1.69 & 60.6 & 66.4 & 6.7 &  3705.02 & 1.15 & 2.274 \\
E & Dust-Masked &  2.9 & 1.55 & 60.5 & 66.4 & 3.1 & 3704.50 & 1.16 & 2.541 \\
F & Dust-Corrected & 1.4 & 1.46 & 60.7 & 66.4 & 5.1 &  3704.77 & 1.14 & 2.096  \\
& & (0.03) & (0.005) & (0.1) & (0.1) & (0.1) & (0.10) & (0.003) & \\
\enddata
\begin{singlespace}
\tablecomments{Best-fit parameter values obtained by fitting thin disk dynamical models to the NGC 4786 and NGC 5193 CO(2$-$1) data cubes. We derive 1$\sigma$ statistical uncertainties for the parameters of fiducial models C and F, based on a Monte Carlo resampling procedure and list them under the results for models C and F. These models have 363 and 3181 degrees of freedom, respectively.  The major axis PA, $\Gamma$, is measured east of north for the receding side of the disk. We found the dynamical center of the fiducial model to be at RA=$12^{\mathrm{h}}54^{\mathrm{m}}32.4115^{\mathrm{s}}$, Dec=$-06^{\circ}51^{\prime}33\farcs{920}$ for NGC 4786 and RA=$13^{\mathrm{h}}31^{\mathrm{m}}53.5289^{\mathrm{s}}$, Dec=$-33^{\circ}14^{\prime}03\farcs{546}$ for NGC 5193. These are within $0\farcs{001}$ of the dynamical centers of the other models. The observed redshift, $z_{\mathrm{obs}}$, is used in our dynamical models as a proxy for the systemic velocity of the disk, $v_{\mathrm{sys}}$, in the barycentric frame via the relation: $v_{\mathrm{sys}} = cz_{\mathrm{obs}}$ and is used to translate the model velocities to observed frequency units.}
\end{singlespace}
\label{tabledynparams}
\end{deluxetable*}

\begin{figure*}[t]
\centering
\includegraphics[width=7.0in]{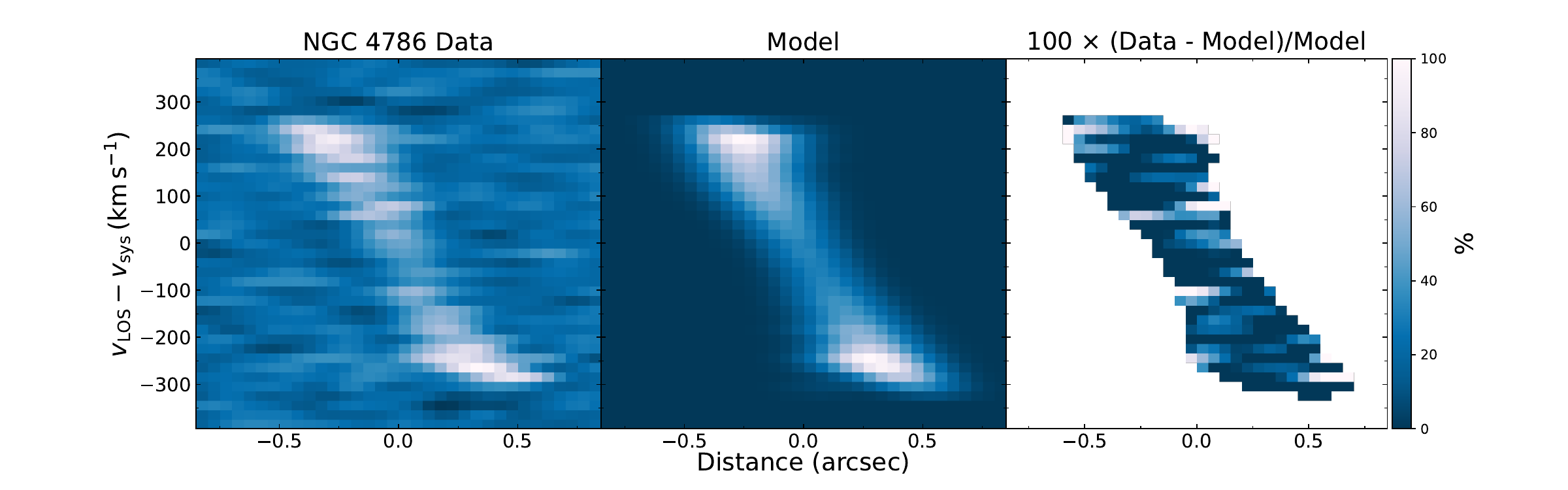}
\includegraphics[width=7.0in]{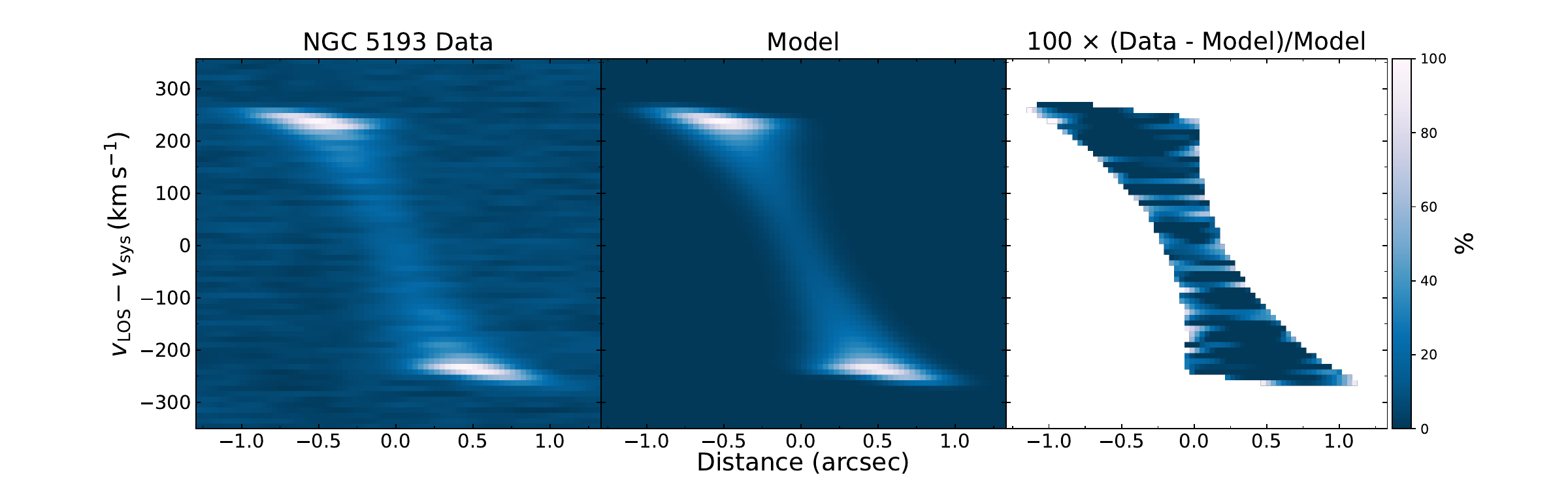}
\caption{PVDs along the major axes of both NGC 4786 (above) and NGC 5193 (below) and their respective best-fit models. Columns show ALMA CO(2$-$1) data (left), models (center), and fractional residuals (right). The PVDs were extracted with a spatial width equivalent to a resolution element for each cube.}
\label{fig:PVDs}
\end{figure*}

\begin{figure*}[t]
    \centering
    \includegraphics[width=6.5in]{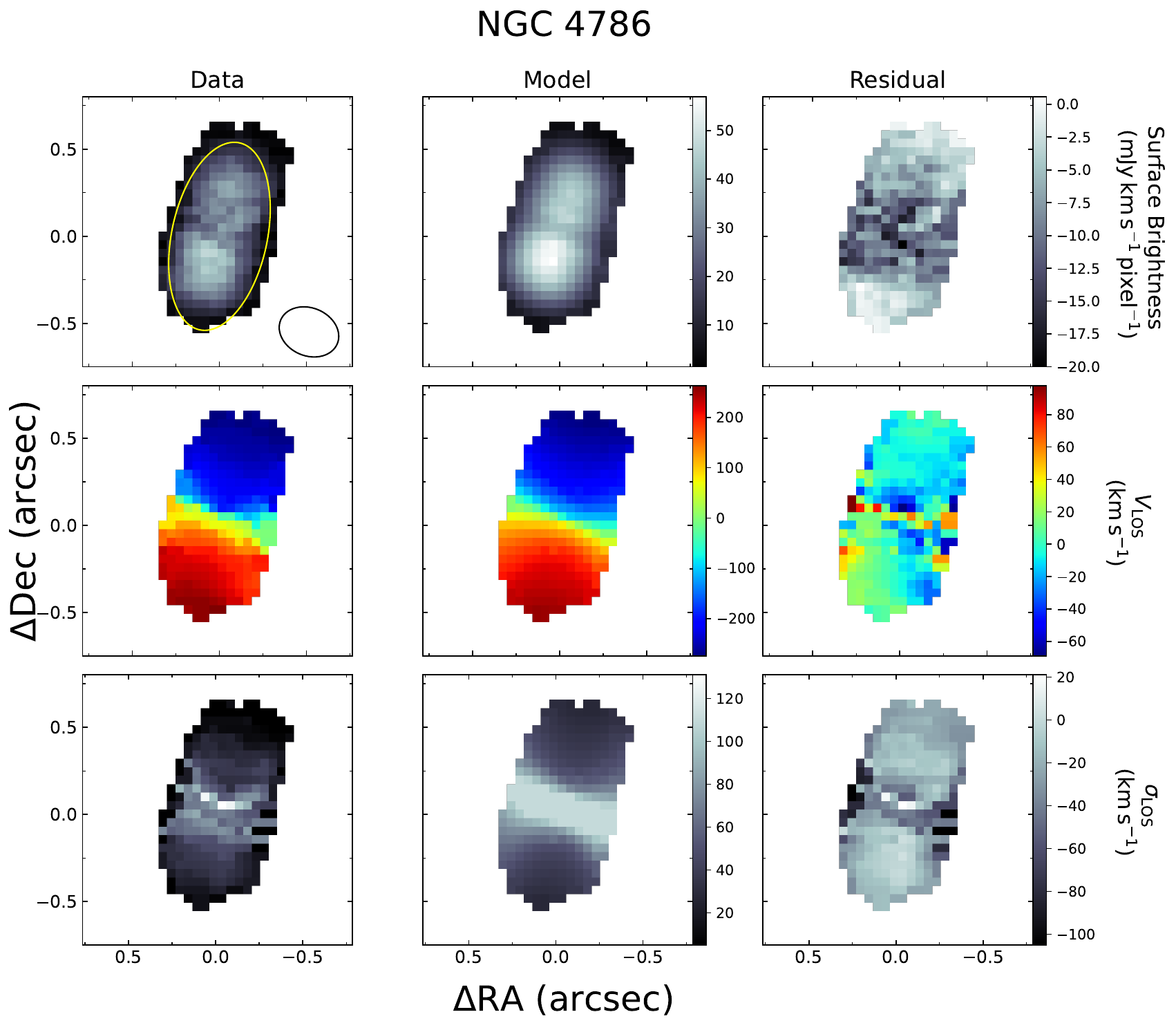}
        \caption{Moment maps for NGC 4786 constructed from the ALMA CO(2$-$1) data cube (left) and its fiducial model (center, model C). Shown are maps of moments 0, 1, and 2, corresponding to surface brightness, LOS velocity $v_{\mathrm{LOS}}$, and LOS velocity dispersion $\sigma_{\mathrm{LOS}}$. The units for the surface brightness map are mJy $\mathrm{km}\,\mathrm{s}^{-1}\,\mathrm{pixel^{-1}}$, and the units for the $v_{\mathrm{LOS}}$ and $\sigma_{\mathrm{LOS}}$ maps are $\mathrm{km}\,\mathrm{s}^{-1}$. The systemic velocity of $4621 \, \mathrm{km\, s^{-1}}$ estimated from our dynamical models has been removed from $v_{\mathrm{LOS}}$. Maps of (data-model) residuals are shown in the rightmost column. The coordinate system is oriented such that $+x$ corresponds to East and $+y$ corresponds to North. While the line profile fits have been determined at each pixel of the full disk, the elliptical fitting region used in calculating $\chi^2$ is denoted in the top left panel with a yellow ellipse. The synthesized beam is represented by an open ellipse in the bottom left corner of the same image.}
    \label{fig:NGC4786Moments}
\end{figure*}

\begin{figure*}[t]
    \centering
    \includegraphics[width=6.5in]{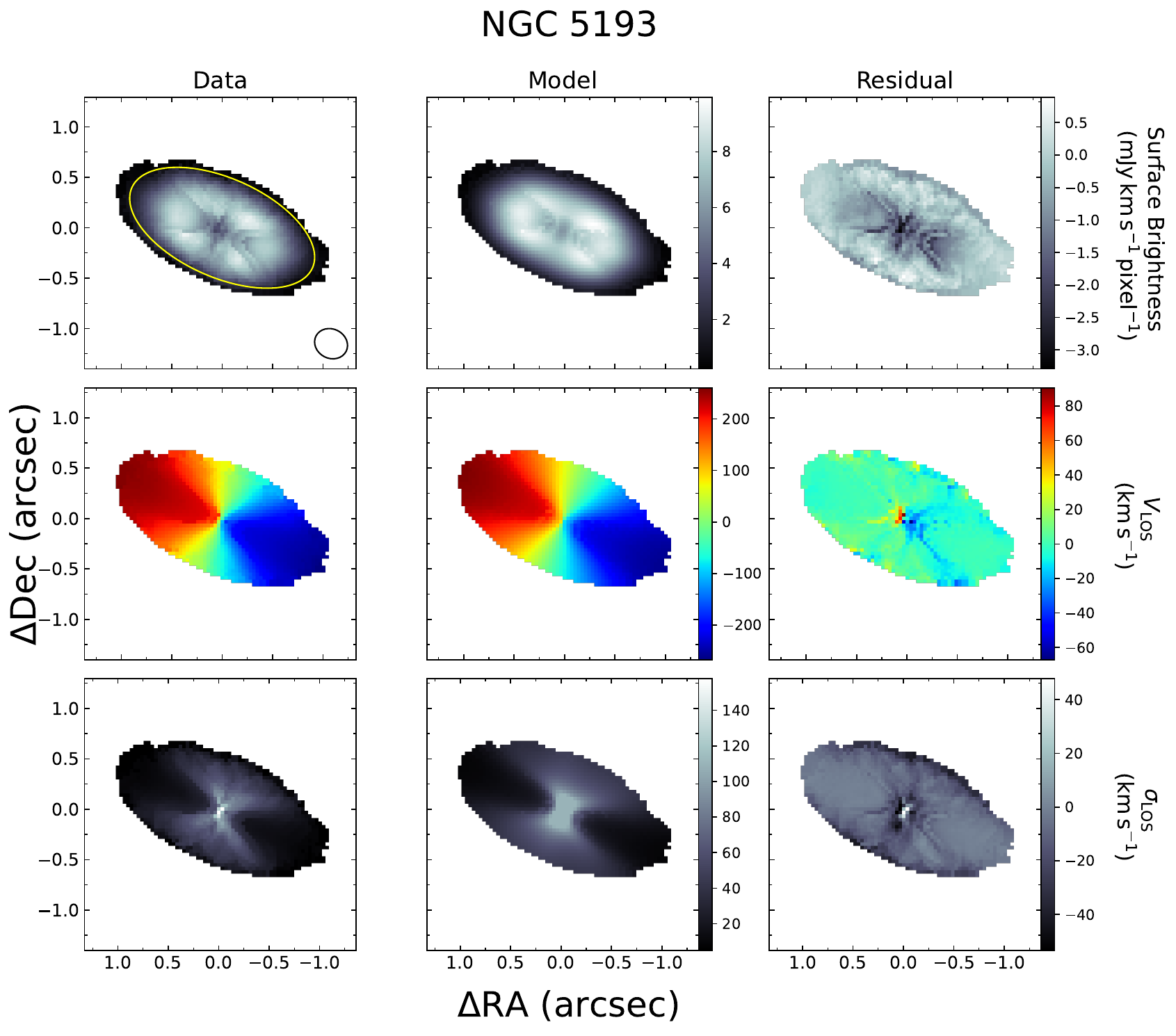}
    \caption{Moment maps for NGC 5193 constructed from the ALMA CO(2$-$1) data cube (left) and its fiducial model (center, model F; see Section \ref{sec:NGC5193Results}). Shown are maps of moments 0, 1, and 2, corresponding to surface brightness, LOS velocity $v_{\mathrm{LOS}}$, and LOS velocity dispersion $\sigma_{\mathrm{LOS}}$. The units for the surface brightness map are mJy $\mathrm{km}\,\mathrm{s}^{-1}\,\mathrm{pixel^{-1}}$, and the units for the $v_{\mathrm{LOS}}$ and $\sigma_{\mathrm{LOS}}$ maps are $\mathrm{km}\,\mathrm{s}^{-1}$. The systemic velocity of $3705 \, \mathrm{km\, s^{-1}}$ estimated from our dynamical models has been removed from $v_{\mathrm{LOS}}$. Maps of (data-model) residuals are shown in the rightmost column. The coordinate system is oriented such that $+x$ corresponds to East and $+y$ corresponds to North. While the line profile fits have been determined at each pixel of the full disk, the elliptical fitting region used in calculating $\chi^2$ is denoted in the top left panel with a yellow ellipse. The synthesized beam is represented by an open ellipse in the bottom left corner of the same image.}
    \label{fig:NGC5193Moments}
\end{figure*}

\begin{figure*}[t]
\centering
\includegraphics[width=7in]{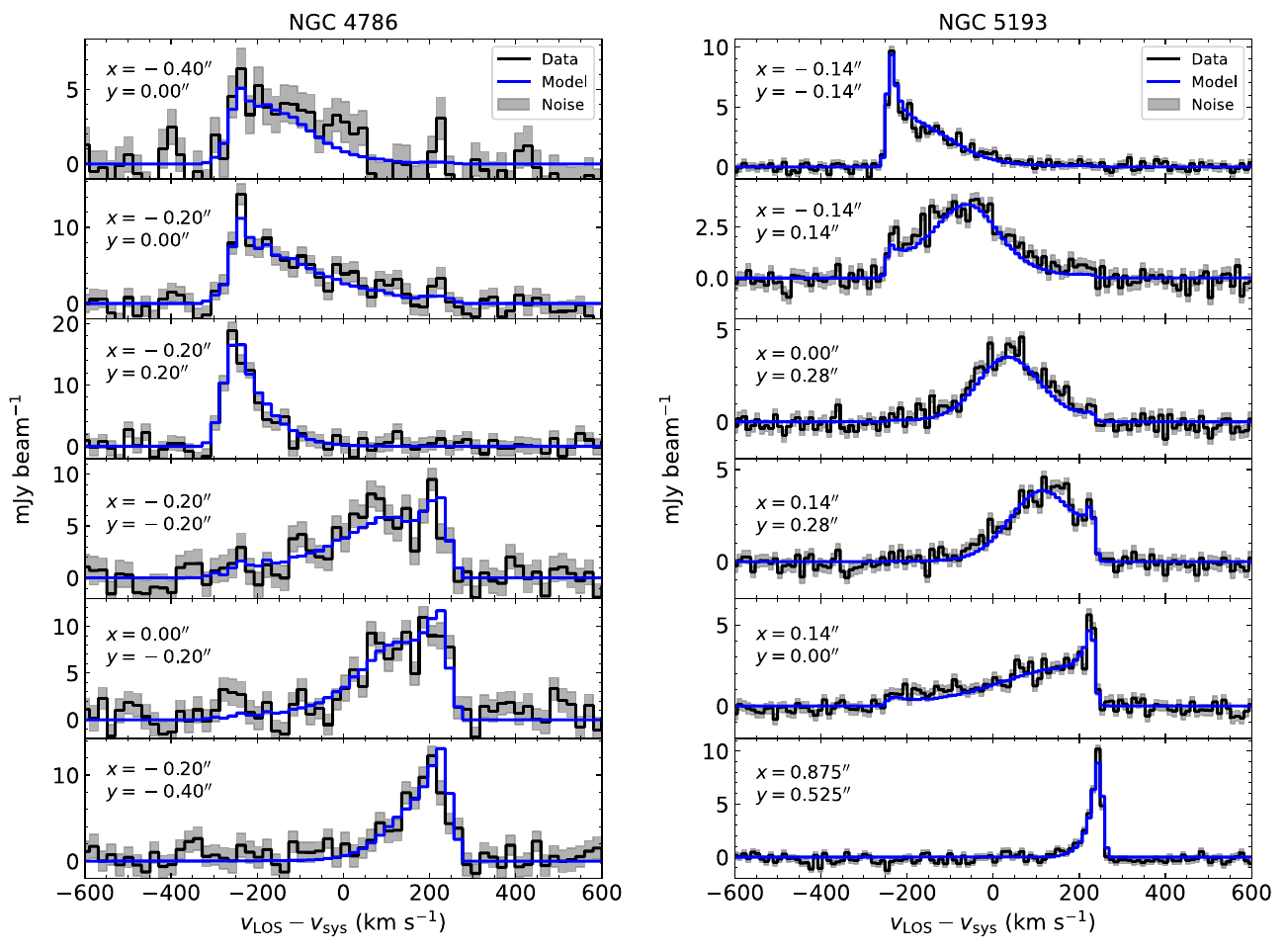}
\caption{CO line profiles extracted from six spatial locations within the block-averaged NGC 4786 and NGC 5193 data cubes, along with their respective fiducial models (models C and F). The $x$ and $y$ positions are given relative to the disk dynamical center, with $+y$ indicating North and $+x$ indicating East. The gray shaded area represents values in the range of data $\pm 1\sigma$, where the $1\sigma$ value is from our 3D noise model used in the $\chi^2$ optimization.}
\label{fig:LineProfiles}
\end{figure*}

\subsection{NGC 4786 Dynamical Modeling Results}
\label{sec:NGC4786Results}
We present the results for three dynamical models (A,B,C) for NGC 4786 in Table \ref{tabledynparams}. The difference among them is the input host galaxy MGE model, which we described in Section \ref{sec:HostGalaxyModeling}.  In summary, dynamical models A-C yield a range in BH mass that spans $(3.9-5.8)  \times 10^8\, M_{\odot}$ with reduced $\chi^2$ ($\chi^2_{\nu}$) values between 1.421 and 1.488 over 363 degrees of freedom. Using the dust-corrected MGE as an input, dynamical model C is the statistically best-fitting model, and we adopt it as our fiducial model to use in our systematic tests of the error budget. With $\chinu = 1.421$, model C is not a formally acceptable fit when considering the degrees of freedom, which should achieve $\chinu \leq 1.125$ when using a significance level of 0.05. The $\chi^2$ values may be high in part due to inherent limitations of deconvolving the CO flux map. However, while changes to the intrinsic flux map may improve $\chi^2$, they are unlikely to significantly affect \mbh\ (e.g., \citealt{2006MmSAI..77..742M,2013ApJ...770...86W,2021ApJ...908...19B}). An in-depth analysis of the systematic uncertainties of the measurement is described in Section \ref{sec:ErrorAnalysis}. We present the major axis PVD, moment maps, and CO line profiles extracted from the data and fiducial model cubes in Figures \ref{fig:PVDs}, \ref{fig:NGC4786Moments}, and \ref{fig:LineProfiles}. The comparisons highlight good overall agreement between the data and model. The model PVD is able to emulate features such as the broad distribution in rotational velocity that is observed within $r \leq 0\farcs{5}$ on both the approaching and receding sides of the disk, as well as the decrease in velocity towards the center. An analysis of the moment maps shows that the model velocity field captures most of the behavior seen in the outer portions of the disk, where discrepancies in LOS velocity are $<{\sim}20$ km\,${\mathrm{s^{-1}}}$, although noticeable disagreement is seen at the center, particularly along the minor axis. As described by \cite{2016ApJ...823...51B}, along an inclined disk's minor axis, the projected distance between the nucleus of a galaxy and a point along the minor axis is compressed by a factor of $\cos(i)$, and poor spatial resolution across the minor axis can lead to pronounced beam smearing. Given the large ALMA beam (FWHM = $0\farcs{31}$ = 93 pc) relative to this small disk, it is unsurprising that beam-smearing effects are most evident in this region.\,The moment 0 map also highlights discrepancies between data and model in the central region, with the moment 0 map indicating a higher model CO surface brightness than what is observed. The CO line profiles can display complex characteristics, but even though the fine details within the broad and asymmetric data line profiles may be missed, our model CO line profiles generally capture their overall shape. 

As for other free parameters in the model, our dynamically determined $\Upsilon_H$ values, especially for model A, are higher than typical $\Upsilon_H$ values ($\Upsilon_H \le 1.30\ M_\odot/L_\odot)$ seen in single stellar population models that assume either a \cite{2001MNRAS.322..231K} or a \cite{2003PASP..115..763C} initial-mass function (IMF) for an old (10-14 Gyr) stellar population with solar metallicity, but our measurement is consistent within systematic uncertainties with models assuming a \citet{1955ApJ...121..161S} IMF ($\Upsilon_H\sim1.51-1.84\ M_\odot/L_\odot$). This is expected, given that galaxies with large $\sigma_\star$ tend to follow Salpeter-like or even heavier IMFs (e.g., \citealt{2013MNRAS.432.1862C,2020ARA&A..58..577S}). This is also consistent with \citet{2024ApJ...961..127M}, who find that the centers of massive ETGs typically have Salpeter-like or heavier mass-to-light ratios.
\citet{2024ApJ...961..127M} also find $\Upsilon$ gradients in the majority of massive ETGs, which can affect the measured \mbh.
However, the $\Upsilon$ gradients are measured at radii of $\sim0.2-1.0$ kpc, beyond the physical scale of the dust disks of the galaxies in this study.

The $\sigma_0$ parameter remains fairly low between $9.3 - 10.8 \, \mathrm{\kms}$, which is less than the data cube's channel spacing and is consistent with other gas-dynamical modeling of ALMA data \citep{2016ApJ...822L..28B,2019ApJ...881...10B,2021ApJ...908...19B,2021ApJ...919...77C,2022ApJ...934..162K}. The flux normalization factor $F_0$ is found to be between $1.54 - 1.56$ in our models and is higher than values seen in previous works, where it is typically closer to unity. Upon inspecting Figure \ref{fig:NGC4786Moments}, a comparison of the data and best-fit model's moment 0 map reveals a noticeable difference in surface brightness, particularly close to the disk's center, where the data appears to have faint CO emission. We explore the systematic effect of the input flux map on the mass measurement's error budget in Section \ref{sec:ErrorAnalysis}.

Based on previous dynamical modeling work, we expect the statistical uncertainty on a given dynamical model fit to be much smaller than the uncertainty associated with the extinction corrections in our host galaxy models. To determine the statistical uncertainty for the NGC 4786 BH mass measurement, we used a Monte Carlo resampling procedure. We generated 100 noise-added realizations of our fiducial model by adding noise to each pixel of the model cube. At each pixel, we drew a randomly sampled value from a Gaussian distribution with a mean of zero and a standard deviation equal to the value of our 3D noise cube at the same pixel. We re-optimized a dynamical model to each of our 100 altered realizations, using the values in Table \ref{tabledynparams} as our initial guesses. We list the standard deviation of each free parameter as the $1\sigma$ uncertainty under the results for model C. For the BH mass, the distribution has a mean of $\mbh = 5.0 \times 10^8\, M_{\odot}$ and a standard deviation of $0.2 \times 10^8\, M_{\odot}$ or approximately 4\% of the mean.

\subsection{NGC 5193 Dynamical Modeling Results}
\label{sec:NGC5193Results}

We present three dynamical models (D,E,F) for the NGC 5193 data cube in Table \ref{tabledynparams}. As in the case of NGC 4786, these dynamical models used three unique host galaxy MGEs that treat the effects of the dust on the stellar light differently. Dynamical models D-F yield a range of $\mbh = (1.4-2.9) \times 10^8\, M_{\odot}$, $\Upsilon_H = 1.46 - 1.69\ M_\odot/L_\odot$, $\sigma_0 = 3.1 - 6.7\, \mathrm{\kms}$, and $F_0 = 1.14 - 1.16$ with $\chi^2_{\mathrm{\nu}} = 2.096 -2.541$ over 3181 degrees of freedom. While our models are generally successful at reproducing the observed kinematics over a majority of the disk, a formally acceptable fit would achieve $\chinu \leq 1.042$ assuming a level of significance of 0.05.
The dynamically determined $\Upsilon_H$ values are consistent with single stellar population models assuming a \citet{1955ApJ...121..161S} IMF.
As with NGC 4786, the $\sigma_0$ values are small and are less than the channel spacing in the data cube. The range in $F_0$ are closer to unity than what was found for NGC 4786, though a lack of central CO emission is noticeable in the data visualizations. The major axis PVD, moment maps, and CO line profiles that compare the data and the best-fit model (model F), are shown in Figures \ref{fig:PVDs}, \ref{fig:NGC5193Moments}, and \ref{fig:LineProfiles}. The PVDs and moment maps show that model F emulates the observed PVD structure and disk kinematics over nearly the full extent of the disk. Similarly to NGC 4786, the most noticeable differences are in the central parts of the disk, as the data appears to be more CO-faint than our model. Additionally, the observed line profiles seen in Figure \ref{fig:LineProfiles} extracted near the center of the disk show complex and asymmetric structure that our model cannot fully describe, as channel-to-channel variations in the amplitudes of the data line profiles are not entirely reproduced, although model F does generally capture their overall shapes well. 

We performed a Monte Carlo simulation to determine the statistical uncertainty of the measurement. As we had done for NGC 4786, we created 100 realizations of model F by adding Gaussian noise. Then, we optimized each realization to produce a new estimate of $\mbh$ and create a distribution. The distribution of $\mbh$ was centered around a mean of $1.4 \times 10^8\, M_{\odot}$, and had a standard deviation of $0.03 \times 10^8\, M_{\odot}$, or 2\% of the mean. We list the standard deviation of all other free parameters determined by this simulation under the results of model F in Table \ref{tabledynparams}.

\subsection{Error Budget of the Mass Measurements}
\label{sec:ErrorAnalysis}
While the statistical uncertainties on $\mbh$ derived from our Monte Carlo procedure are small, there are other sources of uncertainty that stem from different aspects of our model construction. It has been shown by previous dynamical-modeling studies \citep{2019ApJ...881...10B,2021ApJ...908...19B,2021ApJ...919...77C,2022ApJ...934..162K} that the statistical model-fitting uncertainties for a given dynamical model vastly underestimate the total error budget of a given measurement when considering the uncertainties from model systematics. To assess the impact of the systematic uncertainties on the error budget in each galaxy, we took our statistically best-fit dynamical models and performed a number of systematic tests that involved changing aspects of the model construction. We briefly describe and list these changes below: 
\begin{itemize}
\item \textit{Dust extinction}: We explored the impact of dust extinction by creating our three MGE host galaxy models (unmasked, dust-masked, and dust-corrected) described in Sections \ref{sec:DustMaskedUnmaskedMGEs} and \ref{sec:DustCorrectedMGE}. Dust is clearly present at the centers of both systems, and previous gas-dynamical studies \citep{2019ApJ...881...10B,2021ApJ...908...19B,2021ApJ...919...77C,2022ApJ...934..162K} have also shown that even in the best cases where the BH SOI is well-resolved, differences in the assumed host galaxy profiles can lead to large discrepancies in $\mbh$. Therefore, while the intrinsic host galaxy mass profile and the uncertainty in its inner slope may be difficult to ascertain, by building a set of MGE models that account for the presence of dust in different ways, we can effectively produce a set of models that bracket the likely range of profiles.

We adopt model C for NGC 4786 and model F for NGC 5193 as the fiducial model for each galaxy and perform the remaining systematic tests on them for a number of reasons. First, these two models are the statistically best-fitting dynamical models for NGC 4786 and NGC 5193. In addition, previous ALMA BH mass measurements \citep{2019ApJ...881...10B,2021ApJ...908...19B,2021ApJ...919...77C,2022ApJ...934..162K} have shown that the statistical model fitting uncertainties for a given dynamical model vastly underestimate the total error budget of a given measurement when considering uncertainties associated with the host galaxy extinction corrections, and the dust-corrected MGE models used in models C and F are our best estimate of the underlying host galaxy profile.

\item \textit{Radial motion}: Following the approach described by \cite{2022ApJ...934..162K}, we allow for radial motion in our dynamical models, by incorporating a simple toy model which is controlled by a parameter $\alpha$. This parameter controls the balance between purely rotational ($\alpha = 1$) and purely radial inflow ($\alpha = 0$) motion.

\item \textit{Turbulent velocity dispersion}: The gas velocity's dispersion is changed from a spatially uniform term of $\sigma(r) = \sigma_0$ to a Gaussian that is a function of radius: $\sigma(r) = \sigma_0 + \sigma_1\exp[-(r-r_0)^2/2\mu^2]$. This model adds three additional free parameters, with $\sigma_1$, $r_0$, and $\mu$ representing the Gaussian's velocity amplitude, radial offset from $r = 0$, and standard deviation, respectively. While this model is not motivated by any physical mechanism, it allows for more variation in the form of the velocity dispersion, and previous modeling has showed some preference for a moderate increase in $\sigma$ toward the center (e.g., \citealt{2016ApJ...823...51B}).

\item \textit{Fit region}: We adjust the elliptical spatial region used to optimize our dynamical models described in Section \ref{sec:Dynamical Modeling}. The new fitting regions for NGC 4786 and NGC 5193 are ellipses with semimajor axes of approximately 1.25 ALMA resolution elements centered on the disk's dynamical center. With this setup, the models are fit to parts of the disk that are more sensitive to the BH's gravitational potential as opposed to the host galaxy's, but there is now a larger fraction of pixels that are in the central beam-smeared region of the disk. 

\item \textit{Gas mass}: The gas disk's mass contribution to the gravitational potential is removed, and model fits incorporate only the gravitational potential of the BH and the host galaxy. This is done by setting the gas disk's contribution to the total circular velocity to $0 \, \mathrm{\kms}$.

\item \textit{Block-averaging factor}: The process of block-averaging our dynamical models leads to coarser angular resolution, and could potentially limit the models' ability to constrain $\mbh$. To test for this possibility, we optimized a dynamical model where no block-averaging was performed. 
\item \textit{Oversampling factor}: We originally built our models on a grid that was oversampled by a factor of 3 relative to the ALMA data. We set this factor to 1 to test the effect of building our models on a grid that is equal in size to the original ALMA spatial scale. 

\item \textit{Input flux map}: We built a different input flux map to weight the CO line profiles. This flux map was constructed by fitting an azimuthally normalized 3D tilted-ring model  \citep{1974ApJ...193..309R,1989A&A...223...47B} in the \texttt{3DBarolo} program to the ALMA data cubes \citep{2015MNRAS.451.3021D}. The program returns a 3D model data cube from which we can produce a flux map in the same manner as a regular ALMA data cube described in Section \ref{sec:Dynamical Modeling} and use as an input in our dynamical models. 

\end{itemize}

\begin{table*}[ht]
\caption{Systematic Error Tests}
\begin{tabular}{lcccc}
\hline
\hline
\multicolumn{1}{l}{}          & \multicolumn{2}{c}{\textbf{NGC 4786}}                                                                                      & \multicolumn{2}{c}{\textbf{NGC 5193}}                                                                                      \\ \hline
\textbf{Systematic Test}      & \multicolumn{1}{l}{$M_{\mathrm{BH},\mathrm{new}} (10^8 \, M_{\odot}$)} & \multicolumn{1}{l}{$\Delta M_{\mathrm{BH}}$} & \multicolumn{1}{l}{$M_{\mathrm{BH},\mathrm{new}} (10^8 \, M_{\odot}$)} & \multicolumn{1}{l}{$\Delta M_{\mathrm{BH}}$} \\ \hline
Radial motion                 & 5.0                                                                    & $0\%$                                                & 1.4                                                                    & $0\%$                                                 \\
Turbulent velocity dispersion & 5.0                                                                    & $0\%$                                               & 1.5                                                                    & $+7\%$                                              \\
Fit region                    & 4.7                                                                    & $-6\% $                                             & 1.3                                                                    & $-7\%$                                              \\
Gas mass                      & 5.0                                                                    & $0\%$                                                 & 1.3                                                                    & $-7\%$                                              \\
Block-averaging factor        & 5.0                                                                    & $0\%$                                                 & 1.4                                                                    & $0\%$                                                 \\
Oversampling factor           & 4.3                                                                    & $-15\%$                                             & 1.5                                                                    & $+7\%$                                              \\
Input flux map                & 6.2                                                                    & $+21\%$                                             & 1.6                                                                    & $+13\%$         \\
\hline 
\end{tabular}
\tablecomments{Results of the systematic tests performed on the fiducial dynamical model for each galaxy, not including tests that involved changes to the host galaxy MGEs. We list the new value of $\mbh$ that the optimization converged to, as well as the percent change from the BH masses determined for the fiducial models C and F.}
\label{tab:SystematicTests}
\end{table*}

The largest shifts in $\mbh$ are seen in the systematic tests that involve changing the host galaxy model to account for extinction. We describe the changes observed from these tests below and list the results of the other systematic tests that do not involve changes to the host galaxy MGE model in Table \ref{tab:SystematicTests}.

The value of $\mbh$ in both galaxies is highly sensitive to how we account for dust extinction, which is highlighted by the differences in the unmasked, dust-masked, and dust-corrected MGE models shown in Table \ref{tabledynparams}. For NGC 4786, the value of $\mbh$ decreases by about 22\% to $\mbh = 3.9 \times 10^8\, M_{\odot}$ when using the unmasked MGE and increases by 16\% to $\mbh = 5.8 \times 10^8\, M_{\odot}$ when using the dust-masked MGE. In the case of NGC 5193, the dynamical model using the unmasked MGE converged to a BH mass of $\mbh = 1.5 \times 10^8\, M_{\odot}$, only 7\% higher than when using the fiducial model's dust-corrected MGE, but a large factor of 2 increase to $\mbh = 2.9 \times 10^8\, M_{\odot}$ is observed when using the dust-masked MGE. As stated earlier, the dust-corrected MGE models are our best estimate of the intrinsic host galaxy profile.

As a final systematic test, we fixed $\mbh = 0 \, M_{\odot}$ in our dynamical models and reoptimized with the dust-corrected MGE models used in models C and F to see if the models could still reliably emulate the observed gas kinematics without the need for a BH. Again, this resulted in poorer fits to the data, with $\chinu = 1.974$ in NGC 4786 as well as a large increase in $\Upsilon_H$ to 2.60 $M_{\odot}/L_{\odot}$ to compensate for the lack of a supermassive BH. As mentioned earlier, this is a higher $\Upsilon_H$ value than what is seen in single stellar population models. We see a similar result in the dynamical model for NGC 5193, with the model fit yielding a much higher $\chinu = 2.978$ and a higher $\Upsilon_H = 1.58 \, M_{\odot}/L_{\odot}$. The results show that $\mbh = 0\, M_{\odot}$ is highly disfavored for both galaxies. 

As shown in Table \ref{tab:SystematicTests}, most of the systematic tests that did not involve adjustments to the MGE models led to relatively insignificant changes to $\mbh$, but there were a few exceptions. For NGC 4786, the most profound $(> 10\%)$ changes were a 15\% decrease due to the change in oversampling factor and a $21\%$ increase due to the choice of flux map. The models are constructed on an oversampled grid in order to account for potentially steep velocity gradients in the disk, as pixels near the disk center can contain molecular gas spanning a large velocity range. Without any oversampling, these velocity gradients can be be missed, and can subsequently lead to models converging on a less massive BH. As for using the new flux map, our dynamical models not only converged on a higher BH mass, but also substantially different values of $F_0 = 0.93$ and $i = 58.6^{\circ}$ with an improved $\chinu = 1.354$. The change of over $10^{\circ}$ in inclination angle, as well as a 40\% decrease in the flux normalization factor indicate significant differences when modeling the disk's structure. As stated earlier, this flux map was created from a 3D model data cube generated from an automated 3D tilted-ring fit in \texttt{3DBarolo}. The tilted-ring model allows $\Gamma$ and $i$ to be different for each ring, and converged on ring inclinations of approximately $60^{\circ}$, a significant difference from our flat disk models. The differences in the empirically measured and tilted-ring model flux maps can be attributed to the assumptions in the \texttt{3DBarolo} model fit as well as variations in the CO surface brightness across the disk, especially near the disk's center, where the disk is CO-faint. These features are not encapsulated in the azimuthally normalized tilted-ring model, which has a relatively constant surface brightness across the disk. This leads to large differences in the overall flux normalization factor and the inferred disk structure. We attribute the large (13\%) change in $\mbh$ in NGC 5193 when changing its dynamical model's flux map to these reasons  as well.

These tests highlight a clear degeneracy between the BH and stellar mass components in our dynamical models. The systematic uncertainties from the host galaxy modeling dominate over the statistical (${\approx} 2\%$) and the distance uncertainty of ${\approx}15\%$ in NGC 4786 and ${\approx}7\%$ in NGC 5193, as well as the other systematic uncertainties (${\approx} 5 - 20\%$) not associated with the host galaxy models. While the mass range for $\mbh$ found in Table \ref{tabledynparams} is at the ${\approx}20\%$ level in NGC 4786 and the factor of $2$ level in NGC 5193, our dynamical models prefer the presence of a central supermassive BH to reproduce the observed gas kinematics, as opposed to models where no BH is present. 

When considering only the fiducial host galaxy MGE model, and including the uncertainties from the systematic tests in Table \ref{tab:SystematicTests}, the distance to the galaxy, and the statistical fluctuations in the data, the range in $\mbh$ is $(\mbh/10^8\, M_{\odot}) = 5.0 \pm 0.2 \, [\mathrm{1\sigma \,statistical}]\,^{+1.2}_{-0.8} \,[\mathrm{systematic}] \pm 0.75 \,[\mathrm{distance}]$ for NGC 4786 and $(\mbh/10^8\, M_{\odot}) = 1.4 \pm 0.03 \, [\mathrm{1\sigma \,statistical}]\,^{+0.2}_{-0.1} \,[\mathrm{systematic}] \pm 0.1 \,[\mathrm{distance}]$ for NGC 5193 if we add the negative and positive systematic uncertainties from Table \ref{tab:SystematicTests} in quadrature. However, these ranges in $\mbh$ neglect contributions from the uncertainties in the extinction correction of the host galaxy models. 

We will incorporate these systematic uncertainties for a number of reasons. First, while our dust-corrected MGE model is our best estimate of the underlying host galaxy model, we must account for the potentially large variation in the inner slope of the stellar mass profile in each galaxy.  While dust attenuation on the host galaxy light may not appear obvious when comparing the data and model isophotes or surface brightness profiles in Figures \ref{fig:Isophotes} and \ref{fig:SurfaceBrightnessProfiles}, the resulting deprojected host galaxy models produce a broad range in $\mbh$. 

In other works that feature $\mbh$ measurements in ETGs \citep{2017MNRAS.468.4675D,2018MNRAS.473.3818D,2021MNRAS.503.5984S,2023arXiv230406117R}, the surface brightness of the host galaxy has typically been parameterized with only dust-masked MGEs. While masking out the most dust-obscured features in the image prior to fitting an MGE may yield better models than without any masking, it does not fully address the problem of extinction, and limits the model fit to the remaining pixels that are not completely unaffected by dust. As shown in this work and in other previous ALMA $\mbh$ measurements \citep{2019ApJ...881...10B,2021ApJ...908...19B,2021ApJ...919...77C,2022ApJ...934..162K}, uncertainties from the extinction correction far exceed the formal statistical uncertainties, so accounting for a range in the inner stellar mass profile slope and its effect on the inferred $\mbh$ should be explored. Therefore, we strongly advocate that future $\mbh$ studies with ALMA explore this often overlooked component of an $\mbh$ measurement's error budget.

After adding the systematic uncertainties associated with the extinction correction in quadrature with the uncertainties associated with the systematic tests in Table \ref{tab:SystematicTests}, the final measured range in $\mbh$ for NGC 4786 is $\mbh$ is $(\mbh/10^8\, M_{\odot}) = 5.0 \pm 0.2 \,[\mathrm{1\sigma \,statistical}]\,^{+1.4}_{-1.3} \,[\mathrm{systematic}] \pm 0.75 \,[\mathrm{distance}]$ and for NGC 5193 is $(\mbh/10^8\, M_{\odot}) = 1.4 \pm 0.03 \, [\mathrm{1\sigma\,statistical}]\,^{+1.5}_{-0.1} \,[\mathrm{systematic}] \pm 0.1 \,[\mathrm{distance}]$.  We compare our measured ranges of $\mbh$ with predicted values of $\mbh$ from BH-host galaxy scaling relations in the following section. 

\begin{figure*}[ht]
\centering
\includegraphics[width=6.5in]{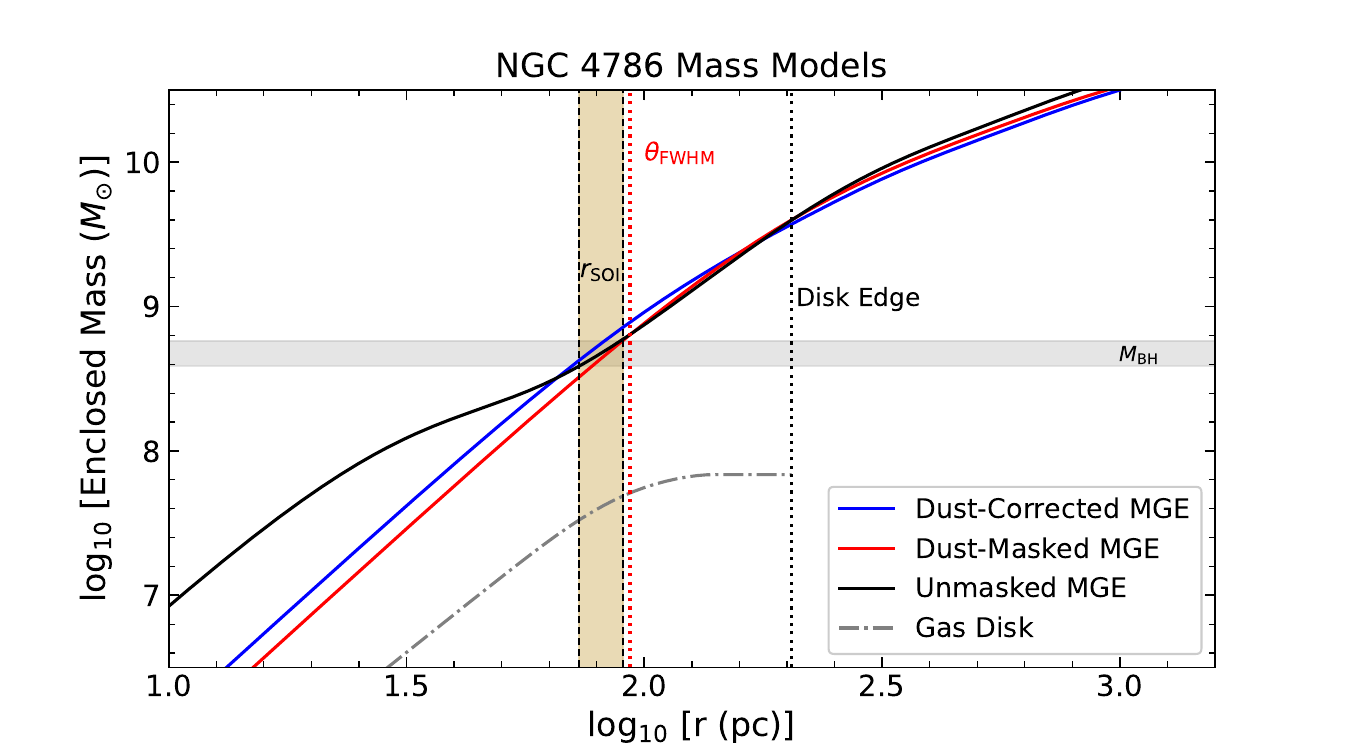}
\includegraphics[width=6.5in]{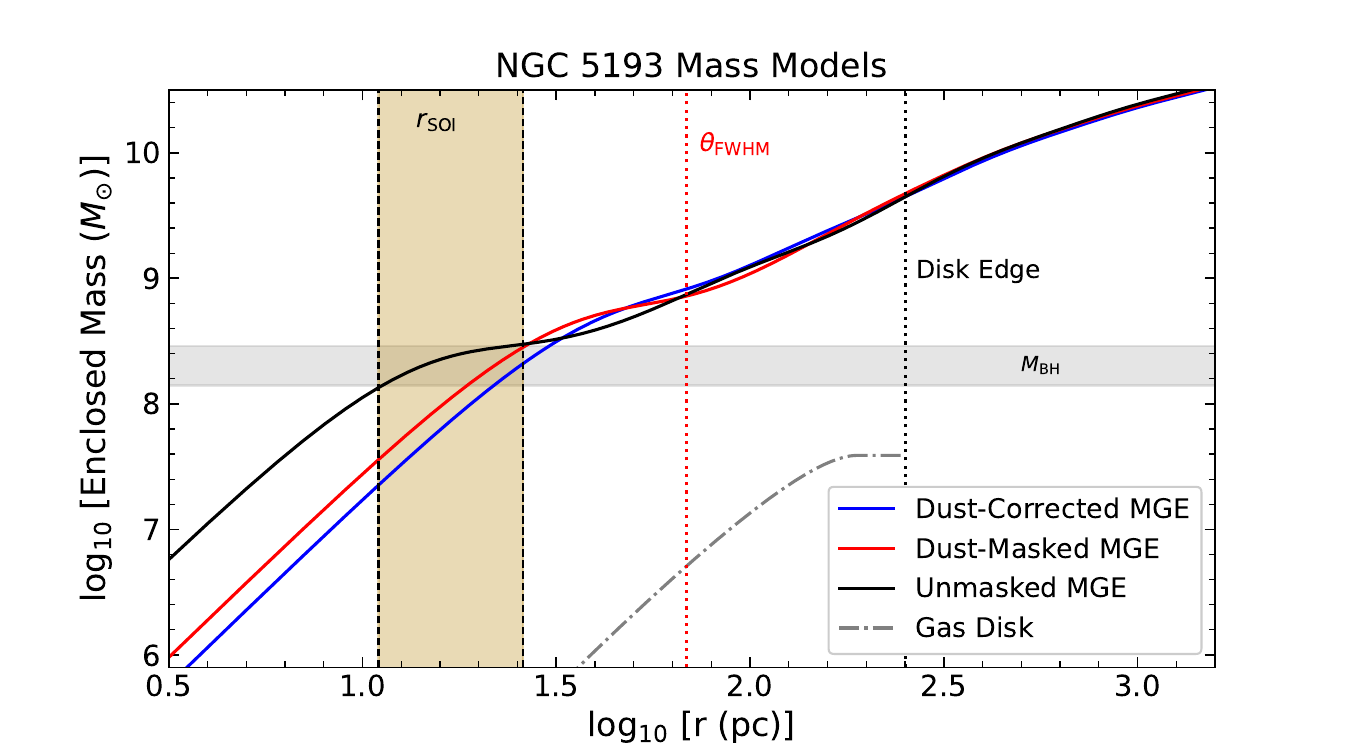}
\caption{Plot of $\log_{10} M_{\star}(r)$ and $\log_{10} M_{\mathrm{gas}}(r)$ vs. $\log_{10} r$ in NGC 4786 (top) and NGC 5193 (bottom) for the three host galaxy MGE models and gas mass profiles used. $M_{\star}(r)$ is calculated via $M_{\star}(r) = rv_{\star,\mathrm{MGE}}^2/G$, where the $v_{\star,\mathrm{MGE}}$ values have been scaled by their respective $\sqrt{\Upsilon_H}$ values listed in Table \ref{tabledynparams}. $M_{\mathrm{gas}}(r)$ is calculated via: $rv_{\mathrm{gas}}^2/G$. The red dotted line indicates the angular resolution of the ALMA observations, whereas the black dotted line represents the outer edge of the dust disk as measured along the major axis. The gray shaded region indicates the range of $\mbh$ determined by our dynamical models using the different input MGE models, while the yellow shaded region enclosed within the black dotted lines indicate the range of the radius of the black hole sphere of influence, $r_{\mathrm{SOI}}$. We have defined $r_{\mathrm{SOI}}$ to be the radius where the stellar and BH masses of a given dynamical model are equal.}
\label{fig:EnclosedMasses}
\end{figure*}

\section{Discussion} 
\label{sec:Discussion}
Our ALMA gas-dynamical mass measurements are the first attempts to measure the central BH masses in NGC 4786 and NGC 5193. In the following subsections, we determine a range for the projected radius of the BH SOI in each galaxy, and describe how using ALMA observations that do not fully resolve this scale leads to large systematic uncertainties in the mass measurement. In addition, we compare our measured ranges of $\mbh$ in each galaxy to predicted ranges from the scaling relations of \cite{2013ARAA..51..511K}.

\subsection{The Impact of Resolving the BH Sphere of Influence in ALMA Gas-dynamical Measurements}
\label{sec:SOI}
Our results indicate that the BH's projected radius of influence, $r_{\mathrm{SOI}}$, is not well resolved in either galaxy, which leads to a degeneracy between stellar and BH mass in our models. We calculated $r_{\mathrm{SOI}}$ in NGC 4786 and NGC 5193 by determining the radius where $M_{\star} = \mbh$ for each of the three input MGE models and highlight the range of $r_{\mathrm{SOI}}$ in each galaxy in Figure \ref{fig:EnclosedMasses}. For NGC 4786, $r_{\mathrm{SOI}}$ is between 73 pc  ($0\farcs{24}$) and  90 pc ($0\farcs{30}$), whereas the ALMA beam FWHM is 93 pc ($0\farcs{31}$), indicating that the SOI is nearly resolved along the major axis of the disk. As pointed out by \cite{2016ApJ...823...51B}, however, the threshold for a successful BH mass measurement rises considerably at higher inclination angles, as projected distances along the minor axis of an inclined disk become compressed by a factor of $\cos(i)$. At an inclination angle of $70^{\circ}$, $r_{\mathrm{SOI}}$ in NGC 4786 is unresolved along the disk's projected minor axis by a factor of nearly 3. As for NGC 5193, we find that $r_{\mathrm{SOI}}$ is between 11 pc ($0\farcs{05}$) and  26 pc ($0\farcs{12}$), and is well below the resolution limit of ${\sim}$69 pc ($0\farcs{31}$).

Following the work of \cite{2013AJ....146...45R}, we compare our measured $r_{\mathrm{SOI}}$ values to the average ALMA beam size through $\xi = 2r_{\mathrm{SOI}}/\theta_{\mathrm{FWHM}}$. For NGC 4786, we find that $\xi = (1.6-1.9$) and for NGC 5193, we find $\xi = (0.3-0.8)$. \cite{2014MNRAS.443..911D} suggests that observations that satisfy $\xi \sim 2$ are adequate for the purpose of conducting molecular gas-dynamical BH mass measurements. This figure of merit was designed to aid in the planning of observational campaigns focused on estimating central BH masses, and is a less stringent threshold than traditional considerations. However, as our results show, measurements based on observations in this regime and lower ($\xi \sim 1$) will have increasingly larger uncertainties. 

While the ALMA observations for NGC 4786 approximately satisfy the aforementioned figure of merit, the sensitivity of $\mbh$ to the choice of host galaxy model in NGC 4786 is still readily apparent. Detecting the enhancement of a tracer particle's circular velocity due to the presence of a supermassive BH is more difficult when the slope of the stellar mass profile in the central region of a galaxy is not well constrained. As we see in Figure \ref{fig:EnclosedMasses}, differences in the overall shape of the different host galaxy mass profiles are minimal at the disk edge and beyond, as expected. However, at radii less than the projected ALMA resolution limit, there are noticeable differences in the overall shape of the stellar mass profiles. The process of converting the observed brightness distributions into mass profiles is complicated by the presence of dust, which limits a dynamical model's ability to reliably separate the BH and galaxy contributions to the overall gravitational potential. Figure \ref{fig:EnclosedMasses} reveals an inner bump in the $M_{\star}(r)$ profile for the unmasked MGE. The bump is unphysical and is a limitation of MGEs that is exacerbated by both the annular extinction of the dust and the constraint that the enclosed mass of each MGE should match at the disk edge. This aspect of MGEs highlights the need for dust-masked and dust-corrected MGEs when modeling the host galaxy's light in cases where dust is readily apparent. 

 In the case of NGC 5193, we see nearly a factor of 2 discrepancy in the measured BH mass from our dynamical models. Quantifying the uncertainty in the inner slope of the NGC 5193 host galaxy mass profile is challenging with the central dust disk, and because the projected BH SOI is unresolved, the range in our measured $\mbh$ is broad. With the BH SOI unresolved by a factor of 2 to 3, the BH mass is a smaller fraction of the total enclosed mass on scales comparable to the observation's resolution limit, and thus, our measurements are even more sensitive to the differences among the stellar mass models. Similarly to NGC 4786, our stellar mass models for NGC 5193 shown in Figure \ref{fig:EnclosedMasses} are well-matched at the edge of the disk, but are noticeably different within the central ALMA resolution element. Moreover, as in the case of NGC 4786, the unmasked MGE displays an unphysical bump at small radii in the $M_{\star}(r)$ profile, which we attribute to the aforementioned reasons provided for NGC 4786.

An additional factor in the measurement uncertainty is the apparent deficit of CO emission in the central region of both galaxies. This deficit is apparent in the moment 0 maps of both NGC 4786 and NGC 5193 shown in Figures \ref{fig:NGC4786Moments} and \ref{fig:NGC5193Moments}, as residuals between the data and best-fit model highlight prominent discrepancies at the center of each disk. Furthermore, the $J-H$ map of NGC 5193 shows additional evidence of a central hole in the circumnuclear disk. In previous work, such as in the case of NGC 6861, these holes can limit the measurement of $\mbh$ to using kinematic information beyond $r_{\mathrm{SOI}}$ and prevent tight constraints on the $\mbh$ \citep{2022ApJ...934..162K}. 
 
\subsection{Comparison to Predictions from BH Scaling Relations} 
The ALMA observations were designed to probe scales comparable to $r_{\mathrm{SOI}}$ using BH mass estimates from the $\mbh - \sigmastar$ relation of \cite{2013ARAA..51..511K} using the definition: $r_{\mathrm{SOI}} = G\mbh/\sigmastar^2$. For NGC 4786, the relation predicts a BH mass of $\mbh = 1.5 \times 10^9\, M_{\odot}$, whereas for NGC 5193, it predicts $3.4 \times 10^8\, M_{\odot}$ when using the $\sigmastar$ values of $286 \,\mathrm{\kms}$ and $205 \,\mathrm{\kms}$ from Hyperleda \citep{2014A&A...570A..13M}. The measured $\mbh$ values for NGC 4786 and NGC 5193 fall below the predicted value, but given the intrinsic scatter of 0.28 dex in the $\mbh - \sigmastar$ relation, the higher end of our $\mbh$ range for NGC 5193 falls within the predicted scatter. \cite{2016Natur.532..340T} suggests that for core galaxies, the core radius $r_{\mathrm{b}}$ is a more robust proxy for $\mbh$ than $\sigmastar$. Given that NGC 4786 exhibits properties of a cored-elliptical galaxy, we input our \texttt{GALFIT} value of $r_{\mathrm{b}} = 0\farcs{46} = 138.5\, \mathrm{pc}$ into the $\mbh - r_{\mathrm{b}}$ (Nuker) relation of \cite{2016Natur.532..340T}. We find that the expected $\mbh$ is $9.7 \times 10^8\, M_{\odot}$. Our mass range of NGC 4786 falls below this relation's predicted value, but within the scatter of the relation.

To compare our mass ranges with the $\mbh - L_{\mathrm{bulge},K}$ relation of \cite{2013ARAA..51..511K}, we used our dust-corrected MGE models in Table \ref{tab:MGE_fiducialcomponents} and assumed a color of $H-K$ = 0.2 mag based on stellar population models of an old stellar population with solar metallicity \citep{2010MNRAS.404.1639V}, as well as absolute solar magnitudes of $M_{\odot,H} = 3.37$ mag and $M_{\odot,K} = 3.27$ mag \citep{2018ApJS..236...47W}. By adding the total luminosity of each component in our $H$-band MGEs and converting them to the $K$-band, we found that $L_{\mathrm{bulge},K} = 3.5 \times 10^{11}\, L_{\odot}$ for NGC 4786 and $L_{\mathrm{bulge},K} =1.3 \times 10^{11}\, L_{\odot}$ for NGC 5193. These yield predicted values of $\mbh$ of $2.5 \times 10^9 \, M_{\odot}$ and $7.4 \times 10^8\, M_{\odot}$ for NGC 4786 and NGC 5193, respectively. Given the scatter of 0.30 dex, our measured BH mass range for both galaxies falls below their respective predicted value and the scatter of the relation.

Finally, we compared our measured ranges with the $\mbh - M_{\mathrm{bulge}}$ relation by converting the total $H$-band luminosities of our dust-corrected MGEs and multiplying them by our $\Upsilon_H$ values for these models listed in Table \ref{tabledynparams}. This was done under the assumption that there are no gradients in $\Upsilon_H$. We found $M_{\mathrm{bulge}} = 5.7 \times 10^{11}\, M_{
\odot}$ for NGC 4786 and $M_{\mathrm{bulge}} = 1.7 \times 10^{11}\, M_{
\odot}$ for NGC 5193. As discussed in \cite{2021ApJ...907....6Z}, a non-neglible fraction of the mass in elliptical galaxies may not belong to the ``bulge" component of the galaxy, and thus, a more nuanced mass decomposition of the host galaxy is needed to reliably calibrate empirical correlations between $\mbh$ and bulge properties. Therefore, our calculated masses should be taken as approximations that likely overestimate $M_{\mathrm{bulge}}$. Indeed, our estimated $M_{\mathrm{bulge}}$ values correspond to predicted values of $\mbh = 3.8 \times 10^9\, M_{\odot}$ and $9.0 \times 10^8\, M_{\odot}$ in NGC 4786 and NGC 5193. Our measured values are below these predictions and the intrinsic 0.28 dex scatter in the relation.

\section{Conclusion}
\label{sec: Conclusion}
We present the first dynamical measurements of the central BH masses in ETGs NGC 4786 and NGC 5193 using ALMA CO(2$-$1) observations that each have $0\farcs{31}$ resolution. In both galaxies, a circumnculear disk in orderly rotation with LOS velocities of ${\sim}270\,\mathrm{km}\,\mathrm{s^{-1}}$ relative to the systemic velocity is observed.

For NGC 4786, our dynamical models constrain the central BH mass to be $(\mbh/10^8\, M_{\odot}) = 5.0 \pm 0.2 \,[\mathrm{1\sigma \,statistical}]\,^{+1.4}_{-1.3} \,[\mathrm{systematic}] \pm 0.75 \,[\mathrm{distance}]$ by fitting the ALMA data with three different host galaxy MGE models. Upon conducting numerous systematic tests, we found that the systematic uncertainties associated with the extinction correction of the host galaxy MGE models were the dominant contributor to the overall error budget. 

In the case of NGC 5193, we fit dynamical models to the ALMA CO(2$-$1) data with three host galaxy MGE models as well. With ALMA observations that resolve on scales of a few BH SOI radii, we found that the uncertainty in the BH mass is around the factor of 2 level. Our measured range of $(\mbh/10^8\, M_{\odot}) = 1.4 \pm 0.03 \,[\mathrm{1\sigma\,statistical}]\,^{+1.5}_{-0.1} \,[\mathrm{systematic}] \pm 0.1 \,[\mathrm{distance}]$ for NGC 5193 highlights the importance of accounting for differences in the host galaxy models due to the presence of dust, as the BH comprises a smaller fraction of the total enclosed mass on scales comparable to the observation's resolution. 

While the mass range for $\mbh$ is broad and our dynamical models do not tightly constrain $\mbh$ in either galaxy, models that contain a central supermassive BH fit the observed ALMA data much better than models without one. Additionally, our models underscore the importance of incorporating a range of plausible inner slopes in the host galaxy mass models when calculating the error budget of an $\mbh$ measurement. This incorporation is necessary when conducting $\mbh$ measurements in dusty systems with ALMA observations that do not fully resolve the projected BH SOI, as stellar mass and BH mass in a dynamical model can become degenerate. Future higher resolution and higher sensitivity observations with ALMA could potentially break the observed BH and stellar mass degeneracy in our dynamical models for NGC 4786 and NGC 5193 and lead to a tighter range on $\mbh$. The higher sensitivity observations would aid in the detection of possible faint CO emission within $r_{\mathrm{SOI}}$. For now, our measurements place important mass constraints on the central BHs in two ETGs that did not have a prior dynamical $\mbh$ measurement, and highlight important limiting factors and considerations for future gas-dynamical $\mbh$ measurements with ALMA. 

\begin{acknowledgements}
This paper makes use of the following ALMA data: {ADS/JAO.ALMA\#2015.1.00878.S} and {ADS/JAO.ALMA\#2017.1.00301.S}. ALMA is a partnership of ESO (representing its member states), NSF (USA) and NINS (Japan), together with NRC (Canada), MOST and ASIAA (Taiwan), and KASI (Republic of Korea), in cooperation with the Republic of Chile. The Joint ALMA Observatory is operated by ESO, AUI/NRAO and NAOJ. The National Radio Astronomy Observatory is a facility of the National Science Foundation operated under cooperative agreement by Associated Universities, Inc. Research at UC Irvine was supported in part by the NSF through award SOSPADA-002 from the NRAO. Support for HST programs \#15226 and \#15909 was provided by NASA through a grant from the Space Telescope Science Institute, which is operated by the Association of Universities for Research in Astronomy, Inc., under NASA contract NAS5-26555. We thank the CAMPARE program for providing research project support for J.M. Sy during the summer of 2020 and for J. Flores-Vel\'azquez during the summer of 2021. JRD thanks the Brigham Young University Department of Physics and Astronomy for their Graduate Assistance Awards. LCH was supported by the National Science Foundation of China (11721303, 11991052, 12011540375, 12233001) and the China Manned Space Project (CMS-CSST-2021-A04, CMS-CSST-2021-A06). KMK would like to thank the lead co-author JHC as well as SCDA for their additional support during the revision phase of this study.

\end{acknowledgements}

%

\vspace{8mm}
\facilities{HST (WFC3), ALMA}


\software{astropy \citep{astropy:2013,astropy:2018}, CASA \citep{2007ApJ...655..144M}, LMFIT \citep{2016ascl.soft06014N}, MgeFit \citep{1994A&A...285..723E,2002MNRAS.333..400C}, \citep{2002AJ....124..266P}, \texttt{photutils} \citep{1987MNRAS.226..747J,larry_bradley_2023_7946442}, Tiny Tim \citep{krist04}, JamPy \citep{2008MNRAS.390...71C}, scikit-image \citep{van2014scikit}}
\vspace{10mm}



\appendix
In Tables \ref{tab:MGE_NGC4786_Unmasked_DustMasked} and \ref{tab:MGE_NGC5193_Unmasked_DustMasked}, we list the components of the unmasked and dust-masked MGEs for NGC 4786 and NGC 5193 described in Section \ref{sec:DustMaskedUnmaskedMGEs}.

\begin{deluxetable*}{ccccccc}[ht]
\tabletypesize{\small}
\tablecaption{NGC 4786 Unmasked and Dust-Masked MGE Components}
\tablewidth{0pt}
\tablehead{
\multicolumn{1}{c}{$k$} &
\colhead{$\log_{10}$ $I_{H, k}$ ($L_{\odot}\, \mathrm{pc}^{-2}$)} &
\colhead{$\sigma_{k}^{\prime}$ (arcsec)} &
\multicolumn{1}{c}{$q_{k}^\prime{}$} & \colhead{$\log_{10}$ $I_{H, k}$ ($L_{\odot}\, \mathrm{pc}^{-2}$)} &
\colhead{$\sigma_{k}^{\prime}$ (arcsec)} & \multicolumn{1}{c}{$q_{k}^\prime{}$}\\[-1.5ex]
\multicolumn{1}{c}{(1)} & \colhead{(2)} & \colhead{(3)} & 
\multicolumn{1}{c}{(4)} & \colhead{(5)} & \colhead{(6)} & 
\multicolumn{1}{c}{(7)}}
\startdata
  & \multicolumn{3}{c}{\bf NGC 4786 (Unmasked MGE)} & \multicolumn{3}{c}{\bf NGC 4786 (Dust-Masked MGE)}\\ \hline 
1 & 4.753  &0.056 & 0.789  & 4.314  &0.321   &0.995   \\
2 &4.540  &0.521  &0.800  & 4.435 &0.569  & 0.730  \\
3 &4.153  &1.278  &0.816  & 4.177 & 1.263 &0.819  \\
4 &3.584  & 2.606  &0.759  & 3.566 &2.731   &0.736   \\
5 &3.463 &4.725  &0.795  & 3.439  & 4.735  & 0.811 \\
6 &2.670  &5.783  &0.902  & 2.592  &  5.713 & 0.886  \\
7 &2.560  &8.842 & 0.690 & 2.592  & 7.899  & 0.690   \\
8 &2.682  &13.350  &0.690  & 2.687 & 12.792  & 0.690    \\
9 &2.204 & 15.762  &0.934  & 2.320 & 14.940 &  0.876\\
10 &2.118  & 23.990 &0.690 & 2.134 & 23.646 & 0.690  \\
11 &1.690  &29.132 &0.952  & 1.769 & 27.513  & 0.967  \\
12 &1.389   &60.476 &0.690   & 1.364 & 59.220  & 0.690  \\
13 &0.269  &47.056  &0.690   & 0.527 & 45.221 & 0.690  \\
14 &1.045   &122.830 &0.982  & 1.099 & 113.697  & 0.963 \\
\enddata
\tablecomments{NGC 4786 unmasked and dust-masked MGE solutions created from fitting MGE models in \texttt{GALFIT} to the $H$-band image. The first column is the component number, the second is the central surface brightness corrected for Galactic extinction and assuming an absolute solar magnitude of $M_{{\odot},H} = 3.37$ mag \citep{2018ApJS..236...47W}, the third is the Gaussian standard deviation along the major axis, and the fourth is the axial ratio. Primes indicate projected quantities.}
\label{tab:MGE_NGC4786_Unmasked_DustMasked}
\end{deluxetable*}

\begin{deluxetable*}{ccccccc}[ht]
\tabletypesize{\small}
\tablecaption{NGC 5193 Unmasked and Dust-Masked MGE Components}
\tablewidth{0pt}
\tablehead{
\multicolumn{1}{c}{$k$} &
\colhead{$\log_{10}$ $I_{H, k}$ ($L_{\odot}\, \mathrm{pc}^{-2}$)} &
\colhead{$\sigma_{k}^{\prime}$ (arcsec)} &
\multicolumn{1}{c}{$q_{k}^\prime{}$} & \colhead{$\log_{10}$ $I_{H, k}$ ($L_{\odot}\, \mathrm{pc}^{-2}$)} &
\colhead{$\sigma_{k}^{\prime}$ (arcsec)} & \multicolumn{1}{c}{$q_{k}^\prime{}$}\\[-1.5ex]
\multicolumn{1}{c}{(1)} & \colhead{(2)} & \colhead{(3)} & 
\multicolumn{1}{c}{(4)} & \colhead{(5)} & \colhead{(6)} & 
\multicolumn{1}{c}{(7)}}
\startdata
  & \multicolumn{3}{c}{\bf NGC 5193 (Unmasked MGE)} & \multicolumn{3}{c}{\bf NGC 5193 (Dust-Masked MGE)}\\ \hline 
1 & 5.813  &0.033 &0.895  & 5.412  &0.078  &0.805 \\
2 & 4.713  &0.218  &0.750  &4.371  &0.435  &0.750  \\
3 & 4.405  &0.848  & 0.756 &4.351 &0.931  &0.752  \\
4 & 4.022 &1.910  &0.814   &3.979  &1.980   &0.809  \\
5 & 3.469 &4.511  &0.750  &3.445  &4.649   &0.750  \\
6 & 3.039 &9.706 &0.842  &3.034  &9.919  &0.851   \\
7 & 2.507 &18.437&0.980  &2.476  &19.507   &0.986    \\
8 & 1.692 &44.349 &0.963 &1.569  &50.580   &0.930    \\
\enddata
\tablecomments{NGC 5193 unmasked and dust-masked MGE solutions created from fitting MGE models in \texttt{GALFIT} to the $H$-band image. The first column is the component number, the second is the central surface brightness corrected for Galactic extinction and assuming an absolute solar magnitude of $M_{{\odot},H} = 3.37$ mag \citep{2018ApJS..236...47W}, the third is the Gaussian standard deviation along the major axis, and the fourth is the axial ratio. Primes indicate projected quantities.}
\label{tab:MGE_NGC5193_Unmasked_DustMasked}
\end{deluxetable*}

\bibliography{paperbibliography}{}
\bibliographystyle{aasjournal}



\end{document}